\documentclass[reprint, superscriptaddress, amsmath, amssymb, aps, pra, floatfix]{revtex4-2}

\usepackage{graphicx} 
\usepackage{dcolumn} 
\usepackage{bm} 
\usepackage[normalem]{ulem} 
\usepackage[colorlinks=true, linkcolor=black, urlcolor=black, citecolor=black]{hyperref} 
\usepackage[margin=0mm]{caption}
\newcommand{\ket}[1]{\left| #1 \right>}

\begin{document}

\title{ Electron qubits surfing on acoustic waves: review of recent progress }

\author{Junliang Wang}
    \altaffiliation{Present address: Silicon Quantum Computing Pty Ltd, University of New South Wales, Sydney, Australia}
    \affiliation{Universit\'e Grenoble Alpes, CNRS, Grenoble INP, Institut N\'eel, F-38000 Grenoble, France}
\author{Hermann Edlbauer}
    \altaffiliation{Present address: Silicon Quantum Computing Pty Ltd, University of New South Wales, Sydney, Australia}
    \affiliation{Universit\'e Grenoble Alpes, CNRS, Grenoble INP, Institut N\'eel, F-38000 Grenoble, France}
\author{Baptiste Jadot}
    \affiliation{Universit\'e Grenoble Alpes, CEA, Leti, Grenoble F-38000, France}
\author{Tristan Meunier}
    \affiliation{Universit\'e Grenoble Alpes, CNRS, Grenoble INP, Institut N\'eel, F-38000 Grenoble, France}
\author{Shintaro Takada}
    \affiliation{National Institute of Advanced Industrial Science and Technology (AIST), National Metrology Institute of Japan (NMIJ), 1-1-1 Umezono, Tsukuba, Ibaraki 305-8563, Japan}
    \affiliation{Department of Physics, Graduate School of Science, Osaka University, Toyonaka 560-0043, Japan}
    \affiliation{Institute for Open and Transdisciplinary Research Initiatives, Osaka University, Suita 565-0871, Japan}
\author{Christopher B\"auerle}
    \affiliation{Universit\'e Grenoble Alpes, CNRS, Grenoble INP, Institut N\'eel, F-38000 Grenoble, France}
\author{Hermann Sellier}
    \altaffiliation{Corresponding author: \href{mailto: hermann.sellier@neel.cnrs.fr}{hermann.sellier@neel.cnrs.fr}}
    \affiliation{Universit\'e Grenoble Alpes, CNRS, Grenoble INP, Institut N\'eel, F-38000 Grenoble, France}

\date{\today}

\begin{abstract}

The displacement of a single electron enables exciting avenues for nanotechnology with vast application potential in quantum metrology, quantum communication and quantum computation. Surface acoustic waves (SAW) have proven itself as a surprisingly useful solution to perform this task over large distance with outstanding precision and reliability. Over the last decade, important milestones have been achieved bringing SAW-driven single-electron transport from first proof-of-principle demonstrations to accurate, highly-controlled implementations, such as coherent spin transport, charge-to-photon conversion, or antibunching of charge states. Beyond the well-established piezoelectric gallium-arsenide platform, first realisations of acousto-electronic transport have also been carried out on the surface of liquid helium that promises unique stability and coherence. In this review article, we aim to keep track of this remarkable progress in SAW-driven transport of electron qubits by explaining these recent achievements from basic principles, with an outlook on follow-up experiments and near-term applications.

\end{abstract}

\maketitle

\section{Introduction} 

Surface-acoustic-wave (SAW) technology is an integral part of modern commercial products such as RFID tags, television tuners, cars, touchscreens and communication and positioning devices. In these applications, SAWs are mainly used as sensors \cite{Mandal2022}, transponders \cite{Plessky2010}, pulse-compression filters or band-pass filters \cite{Morgan2010}. Recently, SAWs are also being discussed for novel applications in optical communications to switch signals in photonic waveguides \cite{Poveda2019}. Besides that, the interaction between SAW and magnetic spin waves has triggered investigations on hybrid systems such as skyrmions \cite{Yokouchi2020, Chen2023} and magnons \cite{Xu_sciadv_2020, Matsumoto2024, Hwang2024}, which might enable further miniaturisation of communication systems \cite{kuss2022}. SAWs are even applied in medicine, where they enable contactless displacement and manipulation of microscopic droplets \cite{Wixforth2004, Ding2013, mi14081543, Delsing2019}. 

SAW technology also finds a growing number of applications in quantum technology -- particularly in solid-state devices \cite{Delsing2019}. SAWs enable for instance the manipulation of rail-encoded photonic qubits \cite{Buhler2022} and hybrid implementations with superconducting qubits \cite{Aref2016, manenti2017circuit, Satzinger2018, bienfait2019phonon, bienfait2020quantum, dumur2021quantum, andersson2022squeezing}. One recent outstanding development from the latter is the novel field of phonon quantum optics \cite{Qiao2023}, where the SAW itself carries quantum information. In this review article, however, we put the focus on a different, but very exciting application of SAWs: the transport of single-electron qubits \cite{Hermelin2011, McNeil2011}. Here, the SAW serves as a transport medium, allowing to physically displace a real particle, together with the information encoded in its quantum state.

Historically, SAW-driven electron transport originated from the field of quantum metrology \cite{wixforth1986, Shilton1996, Janssen2000, Ford2017}. It was anticipated to pave the way towards an extremely stable current source \cite{cunningham1999, cunningham2000} resolving the metrological triangle between resistance $R$, voltage $V$, and current $I$. The central idea is to use the electric potential wave that accompanies a SAW in a piezoelectric device to form a train of quantum dots moving along a transport channel. The wave transports a single electron per potential minimum and drags these single electrons from one reservoir to the other. Such a SAW-driven electron train carries a current $I=e\,f_{\rm SAW}$ that is defined by the product of the electron charge $e$ and the SAW frequency $f_{\rm SAW}$. 

The accuracy achieved with this acousto-electric pumps \cite{cunningham1999, cunningham2000, Janssen2000} was only $10^{-4}$ (100~ppm) at a current of about 500~pA. It was soon outperformed by quantum-dot pumps \cite{Blumenthal2007, Kaneko_2016, Yamahata2016, Ford2017, Kataoka2021, Giblin2023}, which nowadays achieve an accuracy of about $10^{-7}$ (0.1 ppm) at comparable rate. Recently however, the acousto-electric approach regained attention via newly developed SAW generation techniques in the GHz domain \cite{ota2023ondemand}. 

Along with the first experimental demonstrations of SAW-driven quantised currents came theoretical proposals to use the electro-acoustic technology to perform spin-based quantum calculations \cite{Barnes2000}. In addition, SAW-driven electron transport was recognised as an elegant and feasible way to establish coherent links in large-scale spin-qubit architectures \cite{Vandersypen2017}. In this perspective, the electro-acoustic transport technique has been implemented in quantum transport experiments with the ambition to interconnect remote stationary quantum dots \cite{Hermelin2011, McNeil2011} forming the local nodes of a future quantum computer.

The spin coherence of this SAW-driven electron transport has been recently demonstrated in single-shot experiments using the piezoelectric GaAs platform \cite{Jadot2021}. In non-piezoelectric platforms such as silicon or germanium, the deposition of piezoelectric layers should also enable the realisation of SAW-driven spin transport, but experimental investigations with such materials have not been performed so far.

Besides the spin, the charge degree of freedom is also of important interest for quantum-information technology \cite{Bauerle2018, Edlbauer2022}, despite being more fragile. SAW-driven electron transport provides an excellent tool to test the quantum dynamics of flying charge qubits. The transport is slow enough for in-flight gate manipulations, and fast enough to operate below the expected charge decoherence time that is in the order of 10~ns in GaAs \cite{hayashi2003coherent, petta2004manipulation, Petersson2010, stockklauser2017strong}.

As a first step in this direction, single electrons have been transported by SAW through tunnel-coupled wires to investigate the partitioning statistics \cite{Takada2019} and the presence of coherent tunnel oscillations \cite{Ito2021}. Follow-up investigations demonstrated Coulomb-mediated antibunching of two electrons sent simultaneously in the tunnel-coupled wires \cite{Wang2023}. These findings pave the way for applications to coherent two-qubit gates using flying charge qubits, and are likely to impact other flying-qubit implementations \cite{Edlbauer2022} such as levitons \cite{Dubois2013, Dasenbrook2015, Glattli2016, moskalets2016fractionally, Bisognin2019, Assouline2023} or quantum-dot pumps \cite{Fletcher2023, Ubbelohde2023}.

Another topic that is both relevant for flying spin and charge qubits is the transmission of the quantum information to other types of elementary particles such as phonons or photons. The emission of single photons from a train of electrons, that is transported within the potential minima of a SAW train, has been recently successfully demonstrated \cite{Hsiao2020}. But the conversion of quantum states from electron spins surfing on a sound wave into photon polarisation has not been achieved yet.

Electron confinement within the SAW-induced potential was identified as a key aspect for further progress in all of these experiments. Accordingly, the acousto-electric transport technique was recently also tested by employing more sophisticated transducer designs. Instead of forming a long SAW train with a regular transducer to transport a single electron within, a non-uniform design was implemented to perform chirped synthesis of a solitary SAW pulse, enabling single-electron transport on a par with the regular method \cite{Wang2022}. This implementation sets novel perspectives for the precision, synchronisation and scalability of SAW-driven single-electron transport and also revived the acousto-electric transport method for quantum-metrology applications \cite{ota2023ondemand}. Although SAW technology is very well developed \cite{Morgan2010}, only a fraction of the original methods for acousto-electric synthesis have been utilised so far for single-electron transport. Accordingly, we expect there to be plenty of room for improvement.

SAW-driven single-electron transport is also applicable in rather exotic frameworks such as the surface of superfluid helium \cite{byeon2021piezoacoustics}. The platform provides outstanding electron mobility and accordingly unprecedented charge coherence is expected. We anticipate that similar implementations will be executable on the even more pristine surface of solid neon \cite{Zhou2022, Zhou2023}.

This review article keeps up with these major advances in SAW-driven electron transfer that have been achieved over the recent years. In section \ref{sec:single-electron-transport}, we start with an explanation of basic principles and properties of SAW-driven single-shot electron transport. After the basics, we elaborate on coherent SAW transport of electron spin states \cite{Jadot2021} in section \ref{sec:spintransport}, and on in-flight manipulation of charge states \cite{Takada2019,Wang2023} in section \ref{sec:charge-transport}. In section \ref{sec:transducers}, original SAW transducers are discussed that are promising to enhance confinement and thus coherence of the electro-acoustic transport technique \cite{Wang2022}. Section \ref{sec:single-photon-source} focuses on electron to photon conversion \cite{Hsiao2020}. Finally, we look beyond results in GaAs heterostructures and review in section \ref{sec:superfluids} SAW-driven electron transport on the surface of superfluid helium \cite{byeon2021piezoacoustics}.

\section{Single-electron transport between distant quantum dots} 
\label{sec:single-electron-transport}

\begin{figure}[!ht]
\centering
\includegraphics[width=60mm]{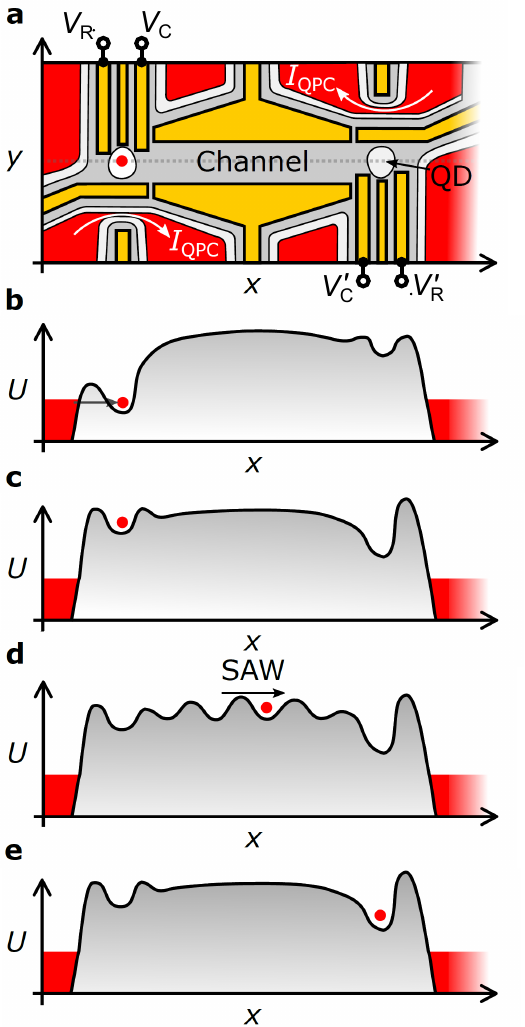}
\captionof{figure}{
    \textbf{SAW-driven single-electron transport.}
    \textbf{(a)} Depleted potential landscape (grey regions) along the transport channel. The surrounding red regions indicate the Fermi sea. The red dot at the source QD (left) indicates an electron. The receiver QD (right) is empty. Next to each QD a quantum point contact (QPC) is placed as electrometer.
    \textbf{(b-e)} Electron potential energy $U$ along the quasi-one-dimensional channel for the following situations: 
    \textbf{(b)} Loading an electron from the Fermi sea.
    \textbf{(c)} Preparation of the isolated electron for SAW-driven transport.
    \textbf{(d)} Transport of the electron by a finite SAW train.
    \textbf{(e)} Catching the flying electron at the receiver QD. 
    Figure reproduced from \cite{edlbauer2019electron} with permission from the author.
    \label{fig:hermelin}
}
\end{figure}

A quantum dot (QD) is a versatile tool to trap and manipulate a single electron. In a GaAs heterostructure, it is typically formed via nanoscale electrodes deposited on the surface that control the electrostatic confinement of the electrons located in the two-dimensional electron gas (2DEG). Highly-controlled SAW-driven single-shot transport of a single electron is achievable \cite{Hermelin2011, McNeil2011} when equipping the ends of a transport channel with such QDs, as sketched in Fig.~\ref{fig:hermelin}a. These two QDs serve as source and receiver for individual electrons, which are transported by the SAW emitted by a piezoelectric transducer (here the SAW is moving from left to right). The presence of an electron within each QD is recorded by quantum point contacts (QPC) which are placed close-by. An electron entering or exiting a QD is detected by tracing the conductance of these very sensitive QPC electrometers.

Let us briefly sketch how SAW-driven single-shot transport of an electron works. Initially, an electron is loaded in the source QD from the close-by reservoir, using fast voltage variations ($\delta V_{R}$ and $\delta V_{C}$) on the QD electrodes -- see Fig.~\ref{fig:hermelin}a and Fig.~\ref{fig:hermelin}b. The loaded electron is then isolated in preparation for transport along the channel -- see Fig.~\ref{fig:hermelin}c. At the same time, the potential of the receiver QD is prepared in anticipation to catch the electron at the end of the depleted transport channel. In this sending configuration, a finite SAW train is launched. As the SAW passes along the channel, its moving potential modulation superposes with the static transverse confinement of the channel and forms a train of quantum dots moving along \cite{Bertrand2016-2}. As a result, the electron loaded in the source QD is picked up by the SAW and gets displaced along the channel -- see Fig.~\ref{fig:hermelin}d. After the SAW has passed the device, the successfully transported electron remains in the receiver QD -- see Fig.~\ref{fig:hermelin}e -- which has been foresightedly prepared in a catching configuration (via the voltage variations $\delta V_{R}^{\prime}$ and $\delta V_{C}^{\prime}$).

\begin{figure}[!ht]
\centering
\includegraphics[width=88mm]{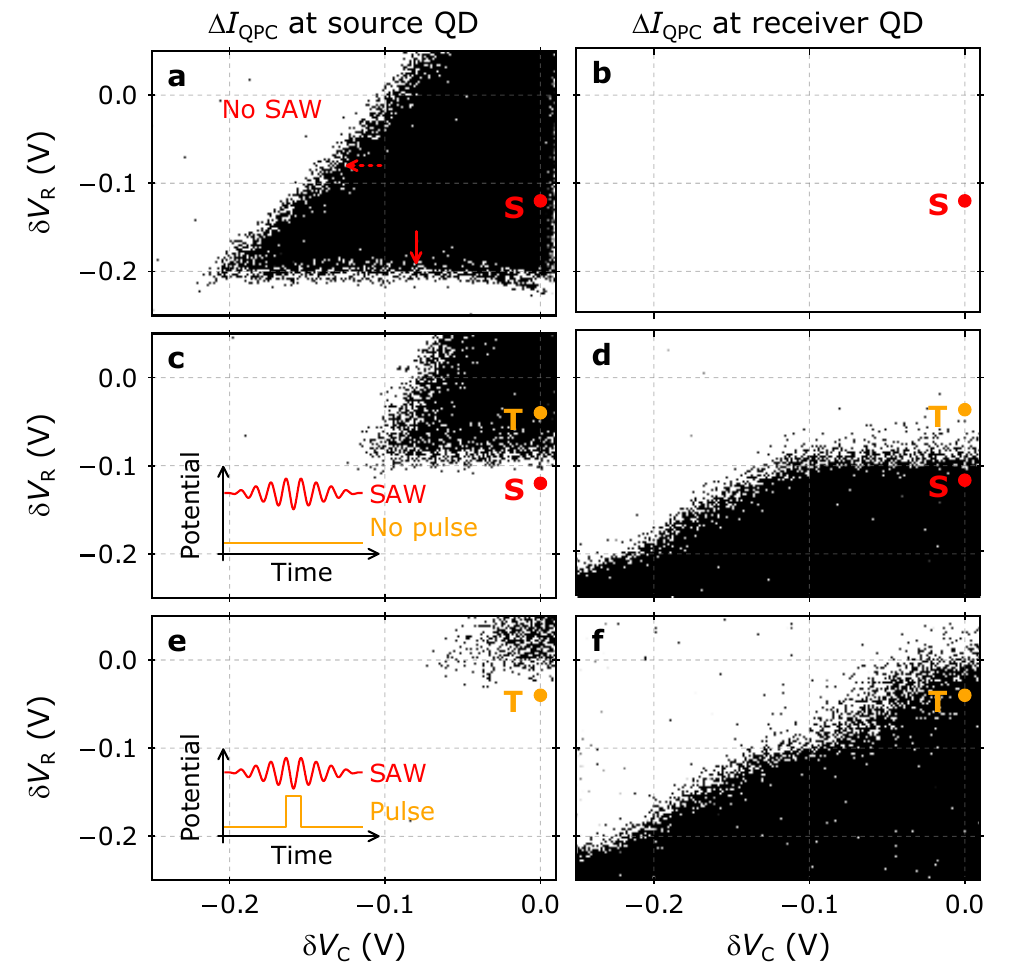}
\captionof{figure}{
    \textbf{Sending and catching maps.}
    QPC current change $\Delta I_{\rm QPC}$ at the source QD (left) and receiver QD (right) as function of the sending configuration on the source QD given by the voltage variations on the gate next to the reservoir ($\delta V_\text{R}$) and next to the transport channel ($\delta V_\text{C}$). The color-map is set such that a black (white) pixel indicates the presence (absence) of an electron.
    \textbf{(a,b)} No SAW is launched during the sequence.
    \textbf{(c,d)} A 30 ns SAW train is launched while the electron is in the sending configuration.
    \textbf{(e,f)} In addition, a short voltage pulse is applied at the plunger gate at the time of the SAW arrival.
    Figure reproduced from \cite{edlbauer2019electron} with permission from the author.
    \label{fig:sendtrig}
}
\end{figure}

During such a transport sequence, the presence of an electron is traced via the QPC currents recorded at the source and receiver QDs. Figure~\ref{fig:sendtrig} shows maps of the difference between the QPC currents before and after the transport sequence, revealing the domain of gate voltages (in the sending configuration) for which the loaded electron is still present in the source QD (black region in the left column) or has been transferred in the receiver QD (black region in the right column). Let us first describe the holding map of the source QD in absence of SAW, for which the electron should not be transferred (black color) -- see Fig.~\ref{fig:sendtrig}a. The black region is limited to the top right part of the map, with two linear thresholds appearing when the electron is lost back either to the reservoir (diagonal threshold with dashed arrow) or into the transport channel (horizontal threshold with solid arrow). On the receiver QD, no change is observed in the charge occupancy (white color), since the receiver QD remains empty -- see Fig.~\ref{fig:sendtrig}b.

Sending now a SAW train through the device, one observes essentially a shift of the horizontal threshold -- see Fig.~\ref{fig:sendtrig}c -- which indicates that the electron in the source QD has been transferred by the SAW into the channel. Verifying the QD occupancy on the receiver side of the transport channel -- see Fig.~\ref{fig:sendtrig}d -- one observes catching events that perfectly correlate with the injection events on the source side. The electron is thus transported from one QD to the other. The timing of the sending event is so far however uncontrolled, and it is also uncertain whether the electron stays in the same SAW minimum (the one in which it was originally captured at the source QD) as it travels through the channel.

If the source QD is now equipped with a surface gate allowing voltage pulses that are faster than the SAW period, one can trigger the sending process for transport in a specific moving potential minimum of the SAW. For this purpose, one brings the electron in a configuration where the SAW alone is not able of taking it away from the source QD -- see configuration T in Fig.~\ref{fig:sendtrig}c,d. In this configuration (T), only the additional application of the voltage pulse activates the sending process -- see Fig.~\ref{fig:sendtrig}e,f.

\begin{figure}[!ht]
\centering
\includegraphics[width=88mm]{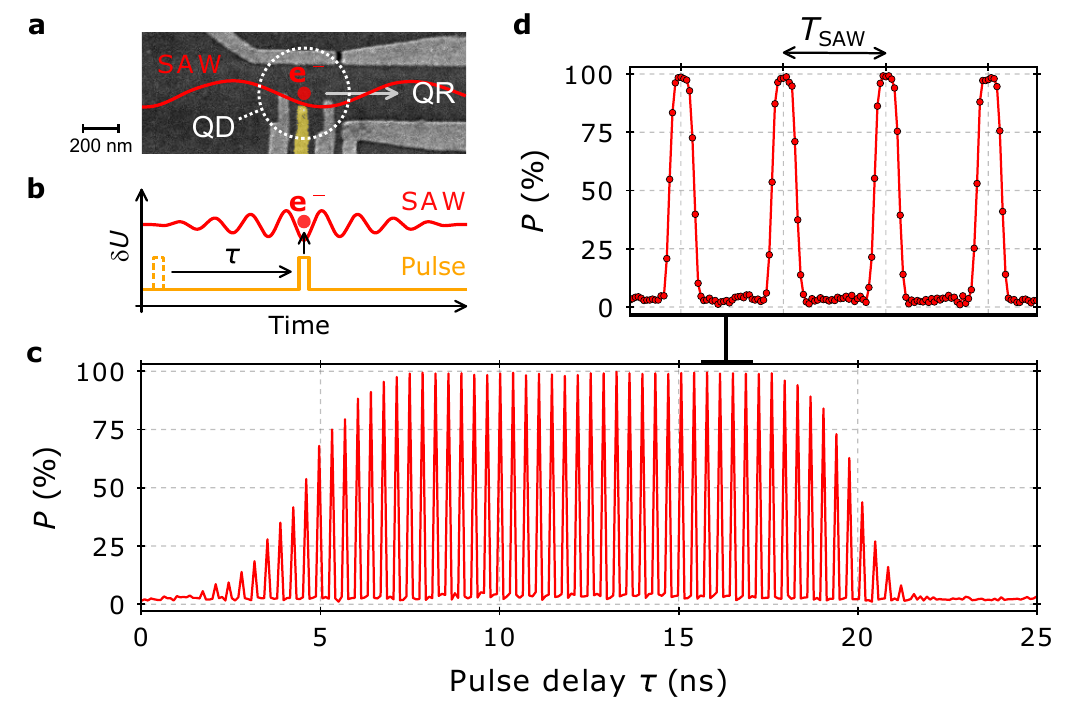}
\captionof{figure}{
    \textbf{Pulse-triggered single-electron transfer.}
    \textbf{(a)} SEM image of the source QD showing the pulsing gate highlighted in yellow.
    \textbf{(b)} Measurement scheme showing the potential modulation $\delta U$ in the QD. The delay $\tau$ of a fast voltage pulse is swept within the arrival window of the SAW at the source QD.
    \textbf{(c)} Measured probability $P$ to transfer a single electron with the SAW from the source QD to the receiver QD as a function of $\tau$.
    \textbf{(d)} Zoom in a time range of four SAW periods $T_\textrm{SAW}$.
    Figure reproduced from \cite{Takada2019} with permission from Springer Nature.
    \label{fig:trigger}
}
\end{figure} 

Figure~\ref{fig:trigger} shows the sending probability for such a situation as function of the delay of the sending pulse \cite{Takada2019}. In the time window of the SAW transit across the QD, distinct peaks are apparent, that are spaced by the SAW periodicity, showing that the voltage pulse allows to address a specific moving potential minimum of the SAW train for electron transport.

\begin{figure}[!ht]
\centering
\includegraphics[width=77mm]{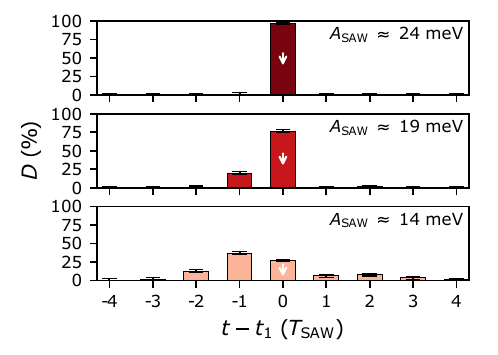}
\captionof{figure}{
    \textbf{In-flight distribution within the SAW train.}
    Distribution $D(t)$ of the electron within the SAW minima for different values of the peak-to-peak SAW amplitude $A_\textrm{SAW}$. $t_1$ indicates the expected arrival time at the barrier gate. The data is obtained via the normalised derivative of transport probability data.
    Figure adapted from \cite{Edlbauer2021} with permission from the American Institute of Physics.
    \label{fig:tof}
}
\end{figure}

At this point, however, it is still not clear if the transported electron stays at the initially addressed position within the SAW train on its journey to the receiver QD. By placing a barrier gate along the transport channel, it is then possible to probe the in-flight distribution of the transported electron within the potential minima of the SAW \cite{Edlbauer2021}. The barrier is set to normally block the passage of the electron, but it can be opened during a fraction of the SAW period using a fast voltage pulse. The time delay of the pulse is then swept over the time window of the SAW arrival at the barrier gate. If the electron is well confined in the targeted SAW minimum, one can expect the transfer probability to rise by 100~\% exactly when the pulse delay coincides with the expected arrival time ($t_1$) of this minimum at the barrier (if the pulse is applied before the electron's arrival, there is no way for the electron to pass, and if it is applied after, the electron is blocked at the barrier until the pulse enables its passage while being in a subsequent part of the SAW). If the electron is located in different minima of the SAW train (for repeated experiments), the transfer probability should rise gradually according to the in-flight distribution of the electron within the SAW train. Figure~\ref{fig:tof} shows an experimental realisation of this time-of-flight measurement. The data demonstrate that the electron is transported in a specific potential minimum of the SAW if the acousto-electric amplitude -- or in other words, the confinement potential within the SAW minima -- is sufficiently high. 

\section{Coherent spin transport} 
\label{sec:spintransport}

A significant academic and industrial effort is underway to build quantum machines based on the control of individual electron spins in silicon quantum processors. Silicon spin qubits are robust with low error, have a small footprint (below 1~µm$^2$) and can be operated at relatively high temperatures (above 500~mK) \cite{Gonzalez2021}. In addition, their close compatibility with the industrial technologies used for the fabrication of classical processors represents a potential path towards integration, for manufacturing and controlling the silicon spin qubits at large scale. In this context, the ability to coherently transfer electron spins within a quantum chip is differentiating the functionalities for solid-state-based qubits. In a different field, displacing atomic qubits has been employed with success to scale atomic quantum processors and it is therefore an important tool to design a realistic quantum processor \cite{Bluvstein2024}. 

As noted by Vandersypen \textit{et al.} \cite{Vandersypen2017}, beyond a certain size of quantum chip and for a finite qubit inhomogeneity within a single quantum device, the fan-out of the gates to control the individual QDs becomes extremely challenging with today’s lithography and fabrication processes. Rather, large-scale quantum structures will have to rely on some form of multiplexing. To make room for this addressing electronic circuitry, one strategy is to separate the quantum core into smaller registers, spatially separated to open enough room for a classical circuitry, which could include multiplexers to reduce the final fan-out wire density. However, that solution implies that all registers can be coherently coupled, on-demand, at least to their nearest neighbors separated by several microns. Mainly two types of coherent couplings have been proposed and experimentally implemented, either by sharing a coherent photon (spin-to-photon conversion in microwave resonators via spin-orbit coupling and hybridisation of the charge state with a cavity mode) \cite{Imamoglu1999, Vandersypen2017, viennot2015coherent, samkharadzeStrongSpinphotonCoupling2018, miCoherentSpinPhoton2018, borjansResonantMicrowavemediatedInteractions2020, Burkard2020, Blais2021, harvey-collardCoherentSpinSpinCoupling2022, Yu2023} or by coherently displacing entangled spins \cite{Kunne2023, Flentje2017, Mortemousque2021, Jadot2021}.

\begin{figure}[!ht]
\centering
\includegraphics[width=86mm]{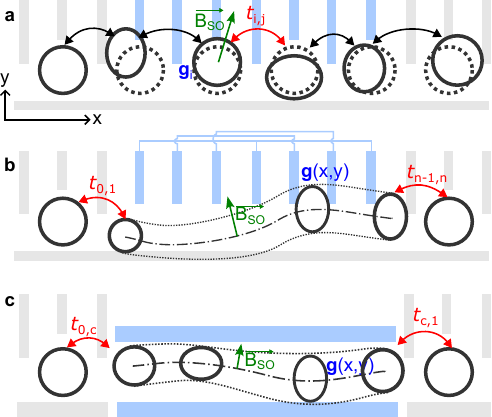}
\captionof{figure}{
    \textbf{Different spin-transfer strategies.}
    \textbf{(a)} Schematic representation of an electron sequentially tunneling from a source dot to a reception dot. The intermediate dots may have small variations in size and shape due to the local electrostatic environment, which in turn may affect the tunneling rates $t_{i,j}$, the electron $\bm{g}$-tensor, or the spin-orbit field $\vec{B}_\textrm{SO}$.
    \textbf{(b,c)} Schematic representation of an electron in a gate-induced (b) and SAW-induced (c) moving potential. These transfer schemes are designed to avoid the successive tunneling events (except at the source and receiver dots).
    \label{fig:spin_transfer_strategies}
}
\end{figure}

The coherent displacement of quantum information within an array (1D or 2D) of QD can be achieved (i) with iterative SWAP operations between spin-exchange-coupled qubits to transfer the spin state over the array \cite{Petta2005}, or (ii) by displacing the charged particle (electron or hole) through the array \cite{Baart2016}. The coherent displacement of a single spin in an array of few QDs was first demonstrated in GaAs/AlGaAs heterostructures, independently in \cite{Flentje2017} and \cite{fujita2017coherent}. Flentje \textit{et al.} performed a series of coherent dot-to-dot tunneling processes in a circular triple QD geometry and probed the coherence of the electron spin state after the displacement. This experiment and the demonstration by Fujita \textit{et al.} opened the way to further investigations in larger arrays \cite{Mortemousque2021}, with holes in germanium QDs \cite{vanriggelendoelman2023coherent} and in academic CMOS devices \cite{yoneda2021coherent}. Recently, coherent spin shuttling over a micron distance has been also performed with electrons in SiGe \cite{Seidler2022}.

There are two challenges to retain spin coherence along a chain of QDs in the so-called bucket-brigade mode. First, the inter-dot tunnel coupling needs to be sufficiently large to permit adiabatic transfer of the electron. Second, the spin interactions -- Zeeman effect governed by the electron Land\'e $\bm{g}$ tensor, spin-orbit (SO) interaction, and hyperfine interaction -- have to be reproducible and constant along the chain of QDs, such that the transfer only induces a deterministic spin evolution -- see Fig.~\ref{fig:spin_transfer_strategies}a.

To avoid the challenge of ensuring an adiabatic spin transfer along a chain of QDs, an alternative is to smoothly displace the confinement potential along the transport direction by changing the voltage excitation on the gates. It has been called the ``conveyor-mode'' approach \cite{xue2023si, Kunne2023, Seidler2022, struck2023spin}. This can be achieved in a less hardware-intensive way by linking the gates of subsequent quantum dots together -- see Fig.~\ref{fig:spin_transfer_strategies}b.

Finally, to reduce the overhead in terms of local electrostatic gates and to minimise the effect of micro-fabrication on the QD properties (shape of the gates, metal granularity), one can also use the confining potential of a SAW to displace a single electron spin with a smooth transfer process at large velocity -- see Fig.~\ref{fig:spin_transfer_strategies}c.

\subsection{Single-spin displacement in a SAW potential}

In section~\ref{sec:single-electron-transport}, we explained how to transfer reliably a single electron between two QDs using a specific minimum of the SAW. In this section, we explain how this technique is applicable to the coherent transfer of a two-electron spin state.

The first demonstration of a single-spin displacement in a moving quantum dot was achieved by Bertrand \textit{et al.} \cite{Bertrand2016}, using the same device as in Hermelin \textit{et al.} \cite{Hermelin2011} which enables the transfer of a single electron through a 4-µm-long channel. Here the spin transfer procedure consisted basically of three steps. At the source dot, an electron spin is first prepared either in $\ket{\uparrow}$ (spin ground state as the electron $g$ factor is -0.4 in GaAs) or in $\ket{\downarrow}$ (spin excited state). Then a SAW burst is launched to pick up the electron and transfer it to the reception dot. There, finally, the spin state is measured.

\begin{figure}[!ht]
\centering
\includegraphics[width=86mm]{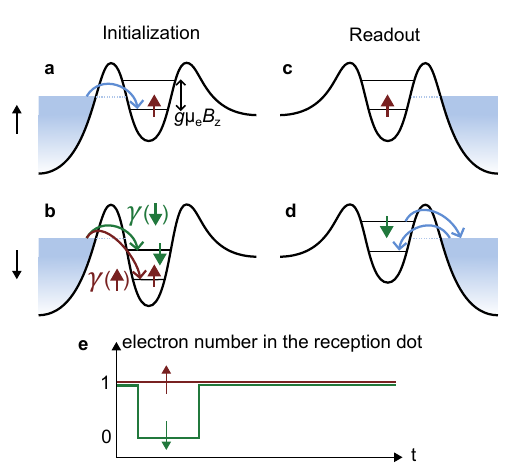}
\captionof{figure}{
    \textbf{Electron spin initialisation and readout.}
    \textbf{(a)} To initialise a single spin in the spin ground state, the chemical potential of the launching dot is set so that only a spin $\ket{\uparrow}$ can tunnel in the dot. 
    \textbf{(b)} By setting the chemical potential of the launching dot below the Fermi sea (for both spin states), the probability of loading either an electron spin up or spin down are assumed equal. At this magnetic field, an equal population of electron spin up and down can tunnel from the electron reservoir. Moreover, the tunnel rates to the QD of an electron spin up $\gamma (\uparrow)$ and down $\gamma (\uparrow)$ are equivalent (similar orbital states for both spin states). Therefore, repeating the single shot experiment, one gets a spin down initialisation statistically equal to $\frac{1}{2}$.
    \textbf{(c,d)} To readout the electron spin, the chemical potential of the reception dot is set so that only an excited spin $\ket{\downarrow}$ can tunnel out to the electron reservoir. Therefore, the readout current $i$ going through a nearby QPC charge sensor would have a temporal signature corresponding to $i(1e) \rightarrow i(0e)$ (electron tunneling out of the QD) followed by $i(0e) \rightarrow i(1e)$ (loading of an electron into the ground state) in case of an electron spin $\ket{\downarrow}$, and $i(1e) \rightarrow i(1e)$ (no electron exchange between the QD and the electron reservoir) for an electron spin $\ket{\uparrow}$, as schematically depicted in \textbf{(e)}. 
    \label{fig:spin_init_readout}
}
\end{figure} 

The initialisation of the electron spin relies on the tunneling process from the reservoir to the quantum dot. An initialisation in the spin ground state can be obtained by setting the potential of the dot such that an electron can tunnel only in the lowest energy state (see Fig.~\ref{fig:spin_init_readout}a). Alternatively, it is also possible to load the electron with the statistical distribution 50~\% $\ket{\downarrow}$ and 50~\% $\ket{\uparrow}$, and wait for the spin relaxation process to occur (see Fig.~\ref{fig:spin_init_readout}b). Then, the procedure for the electron transfer is similar to the one described in section~\ref{sec:single-electron-transport} and repeated over single-shot experiments to obtain statistical information on the spin transfer process. 

The spin readout at the reception dot is based on an energy selective electron tunneling toward the electron reservoir \cite{Elzerman2004}. For that procedure, the chemical potential of the reception dot is tuned such that a spin $\ket{\uparrow}$ state (respectively $\ket{\downarrow}$) has an energy below (above) the Fermi level of the reservoir. Therefore, if the spin of the caught electron is $\ket{\uparrow}$, the electron will stay in the dot (Fig.~\ref{fig:spin_init_readout}c,e). However, if the electron spin state is $\ket{\downarrow}$, the electron will eventually tunnel out of the dot, and a new electron from the reservoir will tunnel in the dot (Fig.~\ref{fig:spin_init_readout}d,e). By recording the charge occupation of the dot with the QPC electrometer, it is therefore possible to determine the spin state of the caught electron.

\begin{figure}[!ht]
\centering
\includegraphics[width=86mm]{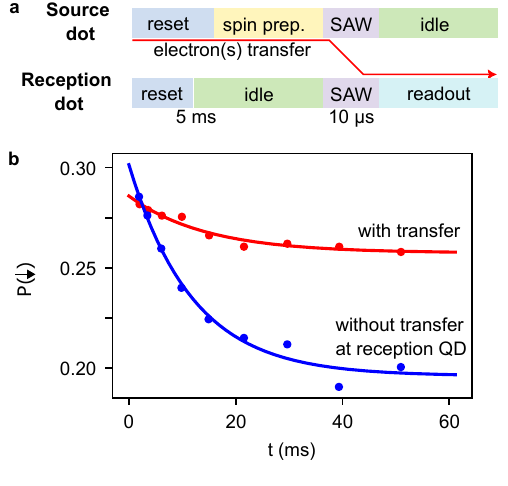}
\captionof{figure}{
    \textbf{Non-local spin relaxation measurements.}
    \textbf{(a)} Schematic sequence used to probe the relaxation of a spin initially prepared in the source dot, then transferred to and measured in the reception dot.
    \textbf{(b)} Probability to measure the electron spin state in $\ket{\downarrow}$ at the reception dot after the SAW transfer from the sending QD as function of the waiting time after initialisation (red). The probability decay amplitude is reduced compared to the calibration experiment (blue), in which the same spin initialisation procedure is performed, but instead of loading into the sending dot and transferring the electron with the SAW, the electron is directly loaded into the reception dot. One can notice the comparable relaxation times in both experiments.
    Figure adapted from \cite{Bertrand2016} with permission from Springer Nature.
    \label{fig:spin_transfer_bertrand}
}
\end{figure}

Figure~\ref{fig:spin_transfer_bertrand} shows the result of an experiment \cite{Bertrand2016} where the spin state is measured at the reception QD after a 50:50 initialisation in the source QD, a waiting time of a few ms, and the electron transfer with the SAW. The spin relaxation shown by the red curve is comparable with that obtained without transferring the electron with the SAW (blue curve), demonstrating the spin conservation during the transfer.

The spin transfer probability in this experiment reached 65~\%. Here the main limitation was the interaction of the SAW burst with the electron during its idling time before and after transfer at the source and reception QDs. Actually, at the source QD location, the SAW-burst electrostatic potential has a finite build-up time (70 SAW periods) due to the IDT design and control scheme. During this period of time, the electron has a lower probability to be injected in a moving potential minimum of the SAW. In addition, the superposition of the static gate-defined electrostatic potential and the moving SAW potential leads to an \emph{oscillating} double quantum dot, which in turn leads to Landau-Zener transitions of the initial spin state in presence of spin-orbit interaction. Similarly, at the reception QD, the electron is subject to the same effect during the SAW burst decay. As we will discuss below, using more appropriate SAW transducers, it is now possible to obtain a single SAW minimum \cite{Wang2022}. Such acousto-electric pulse is expected to significantly enhance the spin transfer fidelity as the perturbation from undesired SAW minima is removed.

In addition to these perturbations at the source and reception QDs, spin relaxation is expected to occur during the electron transport due to spin-orbit (SO) interaction \cite{Huang2013, Zhao2016, Zhao2018}. In a medium without charge noise (gate voltage noise or charge fluctuations in the substrate), the SO interaction would be completely deterministic. The charge transfer would act as a quantum gate on the spin state, which could be measured during a calibration procedure, and thereafter systematically corrected at the reception dot. However, in presence of charge traps such as ionised donors in GaAs/AlGaAs heterostructures -- or charge noise -- the travelling path of the electron may not be reproducible leading to an overall spin relaxation and decoherence.

\subsection{Coherent transfer of a two-electron spin state}

To demonstrate that the electron transfer keeps spin coherence, the evolution of a superposition of spin states during the transfer needs to be investigated. Even though single electron spin superposition is conceptually the simplest, an alternative possibility to create efficiently a superposition of spin states is to use the S-T$_0$ singlet-triplet qubit, defined by two electron spins in a double quantum dot structure \cite{Petta2005}. Its ground state $\ket{0}$ corresponds to the anti-symmetric combination of opposite spins $\ket{S}=(\ket{\uparrow\downarrow}-\ket{\downarrow\uparrow})/\sqrt{2}$. This singlet spin state is energetically favored with respect to the three possible triplet spin states, among which $\ket{T_0}=(\ket{\uparrow\downarrow}+\ket{\downarrow\uparrow})/\sqrt{2}$ is chosen as the excited state $\ket{1}$. This type of qubit exploits the spin exchange interaction between the two particles for qubit control and offers an easy readout procedure (the so-called Pauli spin blockade). More importantly, it represents an efficient way to create superposition and entanglement of spin states when the two electrons are separated. The singlet-triplet qubit is therefore an ideal system for investigations of spin coherence in SAW transport (the coherence time is typically $T_2^*\sim10$~ns in GaAs heterostructures \cite{Petta2005}).

\begin{figure}[!ht]
\centering
\includegraphics[width=82mm]{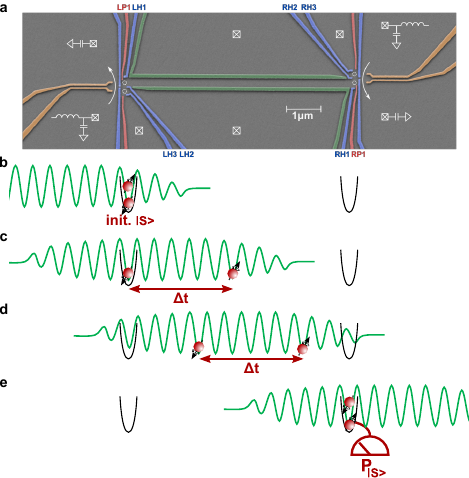}
\captionof{figure}{
    \textbf{Singlet spin transfer scheme.}
    \textbf{(a)} False-color SEM micrograph of the spin transfer device where the transport channel is 6~µm long. Two electrons, initialised in a singlet spin state in a double quantum dot \textbf{(b)}. 
    Subsequently, they are sequentially injected \textbf{(c)} into a moving quantum dot train with a controllable delay $\Delta t$. 
    Once the transfer \textbf{(d)} is complete \textbf{(e)}, the singlet spin probability is measured to determine the spin transfer fidelity. 
    Figure adapted from \cite{Jadot2021} with permission from Springer Nature.
    \label{fig:singlet_transfer_scheme}
}
\end{figure}

Coming back to SAW-assisted shuttling schemes, the ability to separate two electrons initially prepared in a singlet spin state is an elegant way to generate long-range entanglement in quantum processors. Indeed, the two electrons spins would remain entangled even when separated over several micrometers if the shuttling scheme is spin-coherent. A first step towards this coherent quantum link is the demonstration of a coherent singlet spin transfer by a SAW burst, as represented Fig.~\ref{fig:singlet_transfer_scheme}. 

The main challenge to achieve a coherent transfer is the required fine control over the injection/catching processes of each electron from/to a specific moving quantum dot from the SAW burst. Exploiting the deterministic electron triggering scheme presented above, to load in a very precise way the first and second electron into the SAW train, Jadot \textit{et al.} \cite{Jadot2021} demonstrated the ability to shuttle two electrons with a controlled time delay $\Delta t$ varying between 0.5 and 70~ns with a resolution of 0.5~ns. 

\begin{figure}[!ht]
\centering
\includegraphics[width=80mm]{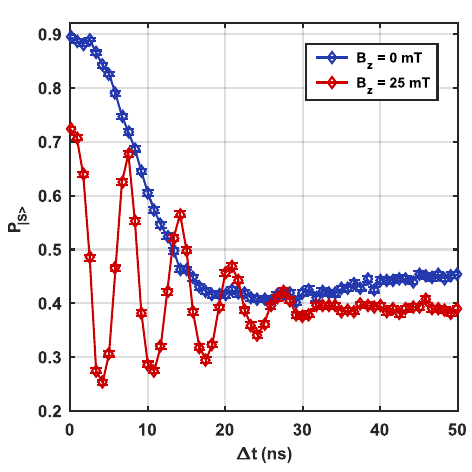}
\captionof{figure}{
    \textbf{Coherent spin transfer demonstration.}
    At zero external magnetic field, the singlet transfer fidelity reaches 89~\% for small separation $\Delta t\approx0$~ns. At $B_z=25$~mT, coherent oscillations appear due to the combination of external field and spin-orbit interaction, with a contrast of $0.567\pm0.007$.
    Figure adapted from \cite{jadot_thesis} with permission from the author.
    \label{fig:singlet_transfer_1D}
}
\end{figure}

Under zero external magnetic field, the authors obtained a singlet transfer fidelity as a function of $\Delta t$ following a Gaussian decay (Fig.~\ref{fig:singlet_transfer_1D}), similar to the case of two electrons separated in a static double quantum dot. A $89.0\pm0.3$~\% maximum fidelity is obtained for $B_z=0$~mT and a minimum injection delay. 

Under a finite magnetic field, oscillations appear in the singlet probability measured after transfer. They are attributed to the spin-orbit interaction affecting each spin during its displacement at the large SAW speed ($v_{\rm SAW}\approx2700$~m/s). They permit to explore coherently the complete Hilbert space of two electron spins, in particular the parallel spin states. The oscillation contrast reaches a maximum of $56.7\pm0.7$~\% under $B_z=25$~mT. As this contrast is above 50~\%, one can conclude that the two electron spins remain entangled even when separated by 6~µm.

\subsection{Spin-orbit effect and decoherence mechanisms}

\begin{figure}[!ht]
\centering
\includegraphics[width=80mm]{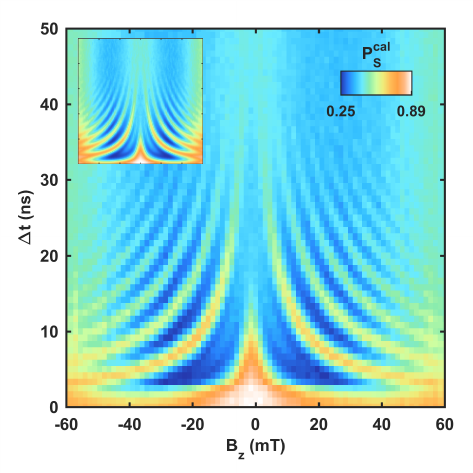}
\captionof{figure}{
    \textbf{High-contrast spin interference pattern.}
    Varying both the external magnetic field and the sending delay, a high-contrast oscillation pattern appears. Each pixel is the average of 10000 realisations. Inset shows a numerical simulation performed using the model described in the main text.
    Figure adapted from \cite{Jadot2021} with permission from Springer Nature.
    \label{fig:singlet_transfer_2D}
}
\end{figure}

Varying the external magnetic field, Jadot \textit{et al.} \cite{Jadot2021} obtained an interference pattern (Fig.~\ref{fig:singlet_transfer_2D}), whose main features are well captured by a simple model based on spin-orbit interaction during shuttling and hyperfine interaction when electrons are idle in the injection or reception dots. During its motion, each electron is submitted to an equivalent magnetic field $\overrightarrow{B_{\rm tot}} = \overrightarrow{B_{\rm ext}} + \overrightarrow{B_{\rm SO}}$, with $\overrightarrow{B_{\rm ext}}$ the applied perpendicular magnetic field and $\overrightarrow{B_{\rm SO}}=22.5$~mT the spin-orbit equivalent magnetic field at the SAW velocity. On the other hand, the hyperfine interaction with the nuclei bath is averaged out during the electron flight (a process called motional narrowing). During its idle time in either the source or reception QD, each spin is however affected by the local magnetic field generated by the surrounding nuclear spins.

This model and the experimental data presented in Fig.~\ref{fig:singlet_transfer_2D} helps to understand the decoherence mechanisms at play during the shuttling protocol. Hyperfine interaction with the sending and catching dot nuclear baths explains the loss of visibility for the large idling times induced by large delays $\Delta t \geq T_2^*$. Charge disorder may also affect the trajectory of the electron, in turn lowering the spin transfer fidelity via the strong spin-orbit coupling. In their result, the authors observed evidence of such disorder-induced decoherence in the reduction of fidelity at $B_z=60$~mT. They propose a mechanism based on the work of Huang \textit{et al.} \cite{Huang2013}, with a $B_z^2$ dependence on the external magnetic field, where the electron trajectory within the depleted channel is perturbed by the random electrostatic background induced by the ionised donors of the AlGaAs layer.

This experiment clearly shows that disorder reduction is necessary to increase the spin-shuttling fidelity. One possible pathway in this regard are undoped heterostructures, using additional surface gates to fill the QD with electrons from a distant ohmic contact \cite{harrellFabricationHighqualityOne1999, See2010, Tracy2014, srinivasanImprovingReproducibilityQuantum2020, ashleaalavaHighElectronMobility2021}. We anticipate that such accumulation gates will also have a beneficial screening effect, making the propagating electron less vulnerable to residual charge fluctuations.

\subsection{SAW-assisted shuttling as a large-scale spin qubit mediator}

SAW-based spin qubit transport is thus validated by experimental realisations, first by Bertrand \textit{et al.} \cite{Bertrand2016} with a single-spin transfer probability of 65~\%, while Jadot \textit{et al.} \cite{Jadot2021} demonstrated a coherent spin transfer protocol of a S-T$_0$ spin qubit. More recently, demonstrations of gate-defined spin shuttling were reported in GaAs/AlGaAs \cite{Baart2016, fujita2017coherent, Flentje2017} and Si/SiGe heterostructures \cite{Zwerver2023, xue2023si}.

\begin{figure}[!ht]
\centering
\includegraphics[width=80mm]{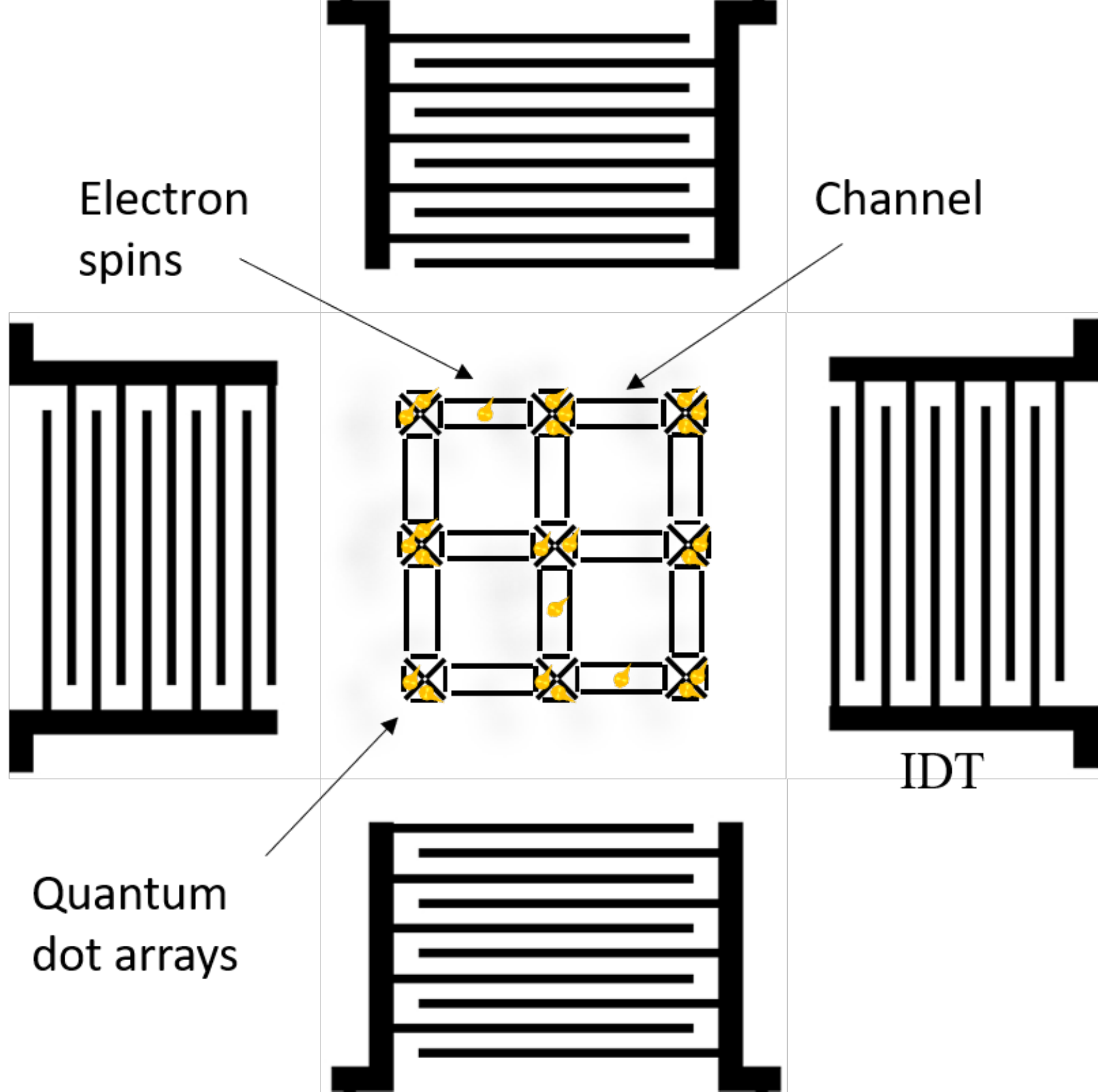}
\captionof{figure}{
    \textbf{SAW-based spin transfer in a quantum processor.}
    Schematics of an interconnected spin-based quantum processor with electron spins transferred through channels between quantum dots using SAW pulses.
    \label{fig:spin_array}
}
\end{figure}

Comparing the two approaches, the first advantage of SAW transfer in comparison with gate-based shuttling is its low overhead both in hardware and signals. Indeed, a single microwave signal is required for the generation of the SAW train, and the long-distance propagation properties of the SAW permit to generate them hundreds of microns away from the quantum core, allowing electron transfer in multiple channels (see Fig.~\ref{fig:spin_array}). The second advantage is its speed, orders of magnitude faster than gate-based shuttling. In particular, this fast motion leads to a strong motional narrowing effect for SAW-based shuttling protocols, while nuclear-spin-free materials are required for a slower gate-based shuttling. However, this high-speed transfer can lead to strong decoherence for materials with non-zero spin-orbit interaction if the electron path is affected by disorder. Similarly, disorder is expected to limit the coherence for gate-defined shuttling protocol, as the shape of the moving quantum dot may evolve during the displacement. Both approaches thus require to minimise electric and magnetic fluctuations along the qubit path, to ensure a smooth reproducible spin qubit evolution during the transfer.

Furthermore, during the SAW-assisted shuttling of a particular electron spin, all the other qubits of the quantum processor should remain localized. The triggered electron injection presented in section~\ref{sec:single-electron-transport} ensures that an electron spin is transported only when it experiences a controlled launch using a fast pulse on the source dot, therefore protecting the other spins from being captured by the SAW.

Finally, implementing SAW-based spin transfer in materials which are not piezoelectric, such as silicon, requires the deposition of a layer of strong piezoelectric material, such as AlN or ZnO, to generate the SAW. Then, the deformation potential of the electron-phonon interaction could be sufficient to couple the mechanical wave to the electron and realize spin transfer.

\section{In-flight operations on charge}
\label{sec:charge-transport}

Having described coherent \emph{spin} transfer as a link between quantum nodes, let us now shift the focus on SAW-driven experiments exploiting the \emph{charge} degree of freedom. In particular, we discuss implementations using a pair of tunnel-coupled transport channels to realise single-electron partitioning and two-electron collision experiments. 

The ability to perform in-flight manipulations of single propagating electrons is a central requirement for quantum-optics-like implementations such as Mach-Zehnder (MZ), Hanbury-Brown and Twiss (HBT) \cite{HBT} and Hong-Ou-Mandel (HOM) \cite{HOM} interferometers. Owing to the rapid progress in semiconductor-device fabrication, pioneering experiments have been realised at a single-electron level in platforms using mesoscopic capacitors \cite{Bocquillon2013}, single-electron pumps \cite{Ubbelohde2014, Fletcher2019, Freise2020}, and levitons \cite{Dubois2013, Jullien2014} (for more details, see the review by B\"auerle \textit{et al.} \cite{Bauerle2018}). These demonstrations paved the way for more advanced single-electron implementations such as the electronic Mach-Zehnder quantum eraser \cite{Kang2007, Weisz2014} or the Elitzur–Vaidman bomb tester \cite{Elitzur_Vaidman_1993}.

In the field of quantum information processing, flying qubits based on photons represent one of the leading quantum computing platforms. Building on the mature field of quantum optics, photonic qubits are commercialised by numerous quantum computing companies such as PsiQuantum \cite{Silverstone2016, alexander2024manufacturable}, Xanadu \cite{madsen2022}, QuiX \cite{Taballione2023} and Quandela \cite{maring2024}. Inspired by the photonic approach, an original proposal is to employ SAW-transported electrons as flying charge qubits for quantum logic implementations \cite{Barnes2000, Bauerle2018}, shifting the well-established paradigm from bosons to fermions.

\begin{figure}[!ht]
\centering
\includegraphics[width=80mm]{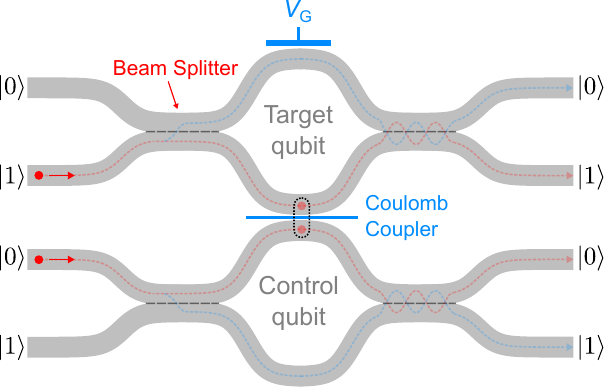}
\captionof{figure}{
    \textbf{A flying qubit based on electron charge.}
    Schematic of two flying charge qubits coupled via a Coulomb coupler that allows to implement a two-qubit gate for flying electrons. Each qubit is composed by a pair of transport paths (grey). The states $\ket{0}$ and $\ket{1}$ are defined by the presence of an electron in the upper and lower channels, respectively. A beam splitter is constructed by coupling both paths with a tunnel barrier (dashed line). The Aharonov-Bohm ring (middle enclosed area) allows to induce a phase shift to the flying electron via a perpendicular magnetic field (not shown) or a side-gate voltage $V_{\rm G}$. The Coulomb-coupling region enables non-linear interaction between a pair of synchronised electrons, resulting in an additional phase shift on the flying electrons. This architecture represents the controlled-phase gate for flying charge qubits.
    \label{fig:flying_qubit}
    }
\end{figure}

Similar to the photonic qubit architecture \cite{Kok2007, OBrien2007}, the electronic quantum state is encoded in the location of the flying qubit within one of two transport paths. As schematically shown in Fig.~\ref{fig:flying_qubit}, the presence of the electron in one or the other path is described as $\ket{0}$ and $\ket{1}$. A superposition state can be prepared in a region where the pair of channels are at close proximity, only separated by a thin tunnel barrier. When the transverse confinement potential is symmetric, the states $\ket{0}$ and $\ket{1}$ hybridise to form the symmetric $\ket{S}=(\ket{0}+\ket{1})/\sqrt{2}$ and the antisymmetric state $\ket{A}=(\ket{0}-\ket{1})/\sqrt{2}$ \cite{Bauerle2018}. As the single electron propagates across this tunnel-coupled region, part of the wavefunction is coherently transmitted to the neighbouring channel, occupying both sides with equal probability. This implementation is the electronic equivalent to the photonic beam splitter, and it will be focus of section~\ref{sec:partioning}.

On the other hand, to induce a phase shift on the flying electron, one can exploit the Aharonov-Bohm effect by separating and then recombining the two transport paths, thereby forming an enclosed surface. The additional phase shift is then controlled by a perpendicular magnetic field or by a side-gate voltage. The first demonstration of such a single qubit rotation has been shown a decade ago using a continuous stream of electrons \cite{Yamamoto2012}. Only recently, this realisation has been achieved at a single-electron level using levitons in graphene \cite{Assouline2023}.

Another essential ingredient for a flying charge qubit is a two-qubit gate, where a target qubit acquires a phase shift in the presence of a control qubit. In contrast to photons which hardly interact with each other, except via their bosonic symmetry relation \cite{Kok2007, OBrien2007, OBrien2009}, electrons offer a direct path to two-qubit gates through their long-range Coulomb interaction \cite{Bertoni2000, Ionicioiu2001, Bauerle2018}. In section~\ref{sec:collision}, we will present recent progress on two-electron experiments which demonstrate the feasibility of this Coulomb-mediated coupling.

\subsection{Partitioning of a flying electron}
\label{sec:partioning}

\begin{figure}[!ht]
\centering
\includegraphics[width=65mm]{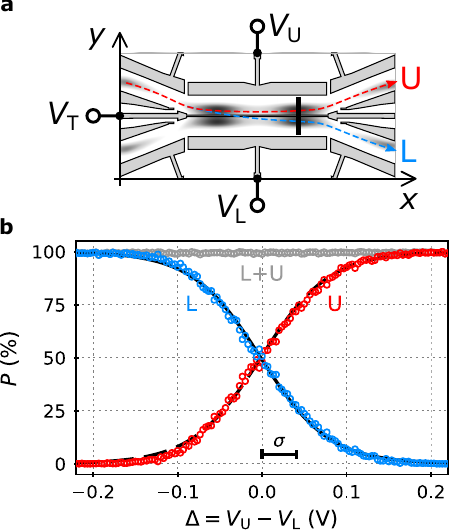}
\captionof{figure}{
    \textbf{Single-electron partitioning.}
    \textbf{(a)} Schematic of a pair of coupled channels for SAW-driven flying electrons. The barrier-gate voltage, $V_{\rm T}$, and the side-gate voltages, $V_{\rm U}$ and $V_{\rm L}$, offer full control of the potential landscape of the coupling region. Dashed arrows indicate a possible trajectory of an electron. The moving confinement SAW minima are schematically indicated by dark elongated spots.
    \textbf{(b)} Probability $P$ of catching the electron at the upper (U) or lower (L) receiver QD for various potential detuning $\Delta$. The transition width is indicated as $\sigma$.
    Figure adapted from \cite{Takada2019} with permission from Springer Nature.
    \label{fig:partition}
}
\end{figure}

The ability to partition a flying electron is an essential ingredient to realise a coherent beam splitter. It was first demonstrated for SAW-driven electron transport by Takada \textit{et al.} \cite{Takada2019}. The authors employed two transport paths that are coupled via a thin tunnel barrier (see Fig.~\ref{fig:partition}a). The side gates, $V_{\rm U}$ and $V_{\rm L}$, and the middle barrier, $V_{\rm T}$, provide fine control of the confinement potential in the coupling region.

In order to study the transfer efficiency along the individual channels, the authors first decoupled the two paths by strongly augmenting the tunnel barrier. Despite the long distance between the source and receiver QDs ($\sim$~22~µm), the authors showed that a single electron is transferable with an efficiency above 99~\%. Such a condition is an essential requirement for high-fidelity flying-qubit operation.

In order to partition a single electron during the flight, the tunnel barrier $V_{\rm T}$ was lowered, and the transmission probability to the neighbouring channel was controlled by applying a detuning voltage $\Delta=V_{\rm U}-V_{\rm L}$ between the side gates. Figure~\ref{fig:partition}b shows the probability of the electron arriving at the upper (U) and lower (L) receiver QD. The data exhibits a gradual change of the transfer probability enabling to set any desired partitioning configuration. In the case of zero detuning, the electron has equal probability to occupy both channels, thus acting as a 50:50 beam splitter. 

To understand quantitatively these partitioning curves, the authors performed numerical simulations of electron transport taking into account the exact geometry and properties of the employed device. Assuming the flying electron to stay in the ground state all along the transfer, the calculations showed that the partitioning transition should spread over a detuning interval of a few µV only. On the contrary, the experiment showed a transition spreading over tens of mV, which was only explainable by assuming excitation of the flying electron to higher energy states. The entrance of the tunnel-coupled region was identified as the main source of excitation, due to an abrupt change by tens of mV in the electrostatic potential, leading to a non-adiabatic evolution of the quantum state. For stronger SAW confinement, the numerical simulations showed that the electron state should become more robust against abrupt potential variations.

To reveal the existence of coherent tunnel oscillations despite the presence of such excitation, it is necessary to improve the precision of the measurements and therefore reduce the statistical fluctuations inherent to single-shot experiments. For this purpose, Ito \textit{et al.} \cite{Ito2021} employed a continuous SAW to drive a steady flow of single electrons through the tunnel-coupled wire and could observe weak tunnel oscillations with a visibility of about 3~\%. In line with the experiment by Takada \textit{et al.}, this low visibility was attributed to the presence of electron excitation. By enhancing the SAW generation techniques, device geometries and materials, it will likely be possible to minimise the non-adiabatic evolution and enable efficient coherent beam splitting of flying electrons with tunnel-coupled channels.

\subsection{Electron pair antibunching}
\label{sec:collision}

To achieve a universal platform of operations with flying charge qubits, besides the preparation of quantum-state superposition with tunnel barriers, it is essential to implement a two-qubit gate. One way to couple two flying charge qubits is to make them interact via their Coulomb interaction. 

\begin{figure}[!ht]
\centering
\includegraphics[width=80mm]{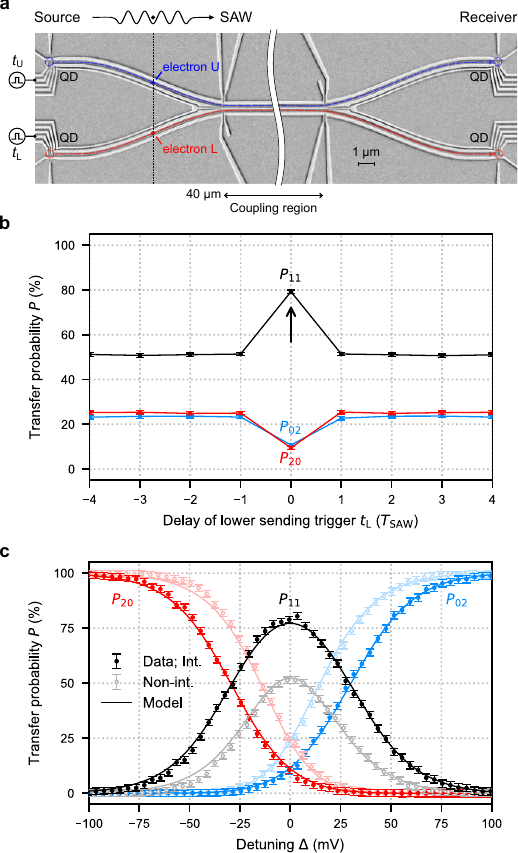}
\captionof{figure}{
    \textbf{Antibunching of an electron pair.}
    \textbf{(a)} Scanning electron micrograph of a SAW-assisted Hong-Ou-Mandel interferometer. A voltage pulse at the upper (lower) source QD with controlled delay $t_{\rm U}$ ($t_{\rm L}$) selectively loads an electron into a desired SAW minimum.  The pair of synchronised electrons (blue and red circles) are transported towards the coupling region before being captured by the receiver QDs.
    \textbf{(b)} Transfer probabilities $P_{20}$ (both electrons at lower detector), $P_{02}$ (both electrons at upper detector) and $P_{11}$ (one at each detector) as function of the time delay between the electrons in units of SAW period $T_{\rm SAW}\approx350$~ps.
    \textbf{(c)} Two-electron partitioning probabilities (data points) when the electron pair is confined in the same (solid color) or different (semi-transparent) SAW minima. The solid lines are predictions from a Bayesian model with and without Coulomb interaction, respectively.
    Figure adapted from \cite{Wang2023} with permission from Springer Nature.
    \label{fig:collision}
}
\end{figure}

Wang {\it et al.} \cite{Wang2023} investigated this electron-electron interaction in a SAW-driven collision experiment using a Hong-Ou-Mandel setup. The authors used a new generation of devices (see Fig.~\ref{fig:collision}a) which integrated three major improvements in order to mitigate electron excitation. First, the transduction efficiency of the interdigital transducer was enhanced by 16~dB by replacing gold with lighter aluminum electrodes \cite{phd_wang}. This larger confinement potential made the transport more robust against potential fluctuations and ensured that the electron remains at the same SAW minimum during the flight \cite{Edlbauer2021}. Second, the gate design was optimised using a quantitative electrostatic model for GaAs heterostructures \cite{Chatzikyriakou2022} in order to reduce the potential variations along the transport path, especially at the entrance of the coupling region. Third, the length of the tunnel-coupled region was extended from 2~µm \cite{Takada2019} to 40~µm \cite{Wang2023} in order to set the propagation time (14~ns) longer than the energy relaxation time ($T_1\sim10$~ns for charge qubits in GaAs heterostructure \cite{Petersson2010}). With such a long coupling region, the flying electrons should have enough time to relax from some excited quantum states towards their ground state.

These improvements enabled the authors to perform a two-electron collision experiment and study the in-flight electron-electron interaction. Two individual electrons were launched simultaneously from two source QDs and the coupling region was tuned to a 50:50 beam-splitter configuration for each electron. The transfer probability was recorded as a function of the time delay between the voltage pulses triggering the electron emission from the two source QDs (see Fig.~\ref{fig:collision}b). When the electrons are travelling in different SAW minima, the probability $P_{11}$ of capturing one electron at each detector shows the classically expected 50~\% probability. Only when the electron pair is confined in the same moving potential, $P_{11}$ increases, reaching up to 80~\%.

This antibunching effect could originate either from the Pauli exclusion of the fermionic statistics, or from the repulsive nature of electron-electron Coulomb interaction. To clarify the origin of the effect, the authors performed this two-electron partitioning experiments as a function of the detuning between the side-gate voltages (see Fig.~\ref{fig:collision}c). They found that the broadening of $P_{11}$ can be explained by an electron-gating effect. Specifically, the presence of an electron on one side of the tunnel barrier modifies the electrostatic potential felt by the other electron. Combining this purely electrostatic effect with the Bayes' theorem, the predicted partitioning probabilities reproduced the experimental data without any adjustable parameter. Therefore, the authors concluded that the antibunching effect is a result of long-range Coulomb repulsion, rather than a quantum-statistical effect.

The strength of the Coulomb interaction was estimated theoretically to be $U_{ee} \approx 0.5$~meV. With such a large interaction, a phase shift of $\pi$ between the electron pair -- the requirement for Bell state preparation -- would be obtained after an interaction time of only 4~ps (equivalent to a propagation length of 12~nm by considering a SAW speed of about 3000~m/s). Owing to such a strong and long-range interaction, the transport channels should be placed further apart, with a wider barrier gate, relaxing the nanofabrication overhead.

It is worth to note that the same electron-gating effect was reported in similar HOM experiments but using single electrons launched into quantum Hall edge channels \cite{Fletcher2023, Ubbelohde2023} in agreement with theoretical studies \cite{Ryu2022, Pavlovska2023}. These experiments therefore demonstrate the feasibility of the Coulomb-mediated coupler as a two-qubit gate implementation for flying charge qubits. For completeness, a similar antibunching effect due to Coulomb interaction has also been observed recently in a very different type of experiment using free electron beams \cite{Haindl2023}.

\subsection{Outlook on in-flight manipulations with real-time control}
\label{sec:inflight-manipulation}

A specificity of SAW-assisted electron transport is its relatively slow propagation velocity of about $3\times10^3$~m/s compared to photons ($3\times10^8$~m/s) and other transport mechanisms of electrons in solid-state devices ($10^4-10^5$~m/s \cite{Kamata2010, Kamata2014, Kataoka2016, Roussely2018, Freise2020}). Owing to this orders-of-magnitude longer timescale and the advances of radio-frequency electronics \cite{THz_electronics}, the SAW platform offers the unique opportunity to dynamically control the electron state during the flight \textit{in real time}. In-flight dynamical control can be implemented by applying tailored voltage waveforms on the tunnel barrier. Synchronising the trigger-send pulse with such a waveform, the barrier height can be arbitrarily controlled along the entire coupling region. 

Regarding the necessary time resolution, Wang \textit{et al.} \cite{Wang2023} showed high-fidelity electron transport over a 40-µm-long coupling region, corresponding to a flight time of 14~ns. Since state-of-the-art arbitrary waveform generators can reach nominal bandwidths above 20~GHz \cite{THz_electronics}, time resolution is not a limiting factor for performing real-time control.

Let us now describe a typical in-flight manipulation of a single flying electron. Before sending the electron, the energy barrier of the middle electrostatic gate is set high to decouple the two channels. When the flying electron is in the coupling region, the barrier is lowered during a short period of time to allow coherent tunneling to the other side. Detecting the statistics of single-shot events at the receiver QDs as a function of the coupling duration, one should observe coherent in-flight Rabi oscillations. This experiment would demonstrate a single-qubit rotation for SAW-assisted flying electrons. In general, dynamical control opens up a plethora of experiments for studying, for example, electron relaxation during the flight, Landau-Zener transitions at the beam splitter or electrostatic inhomogeneities along the channels.

\section{Novel transducers for SAW engineering}
\label{sec:transducers}

Experiments from the last decade showed that the field of flying electrons based on SAW transport has advanced significantly. The main focus was on improving gate geometries and mastering on-demand sending of an electron. However, since the pioneering single-electron shuttling experiments \cite{Hermelin2011, McNeil2011}, the IDT design has barely changed. In this section, we focus on recent developments in IDT designs, and their benefits for SAW-mediated electron transport.

\subsection{Optimisation of transduction efficiency}
\label{sec:saw_ampl}

As mentioned at the beginning of this review, SAW technology has been widely used for quantum applications \cite{Delsing2019}. Numerous innovations have been introduced to improve the transduction efficiency and generate stronger acoustic waves. These innovations include focusing the acoustic beam on a small region \cite{Lima2003, Msall2020}, pulse compression techniques \cite{Schulein2015, Wang2022} and unidirectional SAW emission \cite{Morgan2010, Ekstrom2017, Dumur2019}. Using the latter IDT design for instance, Qiao \textit{et al.} \cite{Qiao2023} were able to strongly couple superconducting qubits with single phonons, demonstrating for the first time the bunching effect between a pair of acoustic phonons.

\begin{figure}[!ht]
\centering
\includegraphics[width=70mm]{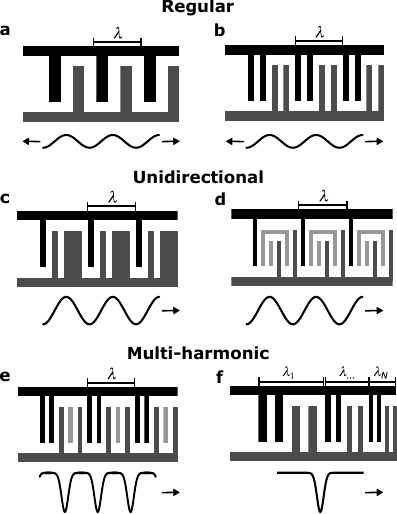}
\captionof{figure}{
    \textbf{Interdigital transducer designs.} 
    Classifications of IDTs with schematic indications on the generated SAW shape (line) and its travelling direction (arrow). 
    Regular or bidirectional IDTs with
    \textbf{(a)} single-finger, and 
    \textbf{(b)} double-finger designs.
    Unidirectional IDTs known as
    \textbf{(c)} DART (distributed acoustic reflection transducer) \cite{Dumur2019}, and
    \textbf{(d)} FEUDT (floating electrode unidirectional transducer) \cite{Ekstrom2017}.
    Transducer designs with multiple resonance frequencies for Fourier synthesis:
    \textbf{(e)} Split-52 IDT \cite{Schulein2015} generating periodic delta-like SAW pulses, and
    \textbf{(f)} chirp IDT \cite{Wang2022} with broadband response producing a single-cycle SAW pulse.
    \label{fig:idts}
}
\end{figure}

Similarly, stronger acoustic waves are also essential for high-fidelity single-electron transfer, by producing stronger confinement in the SAW potential. Therefore, enhancing the transduction efficiency has a direct benefit for SAW-driven electron transport experiments. The transducer employed in the pioneering experiment by Hermelin \textit{et al.} \cite{Hermelin2011} was a regular IDT with a single-finger design (see Fig.~\ref{fig:idts}a). In such a structure, metallic electrodes are located uniformly every half-period $\lambda/2$. Since each electrode acts not only as a SAW emitter, but also as a reflecting mirror, such a design exhibits a cavity effect where SAW is trapped in the transducer for a certain time before leaving the structure \cite{Morgan2010}. Owing to this cavity effect and the low electromechanical coupling coefficient $k^2$ of GaAs, the electron transfer probability was limited to 92~\%.

A double-finger design mitigates the formation of standing waves inside the transducer \cite{Morgan2010} (see Fig.~\ref{fig:idts}b). In the first experimental implementation of this design for SAW-assisted transport, Takada \textit{et al.} \cite{Takada2019} reported a transfer efficiency of 99.7~\%, even though the transfer distance was 5 times longer. Despite this achievement, the SAW confinement potential was not strong enough to keep the electron during transport within the SAW minimum where it was originally loaded \cite{edlbauer2019electron}.

The transduction efficiency can be further enhanced by reducing the mass-loading effect \cite{Morgan2010}. In particular, Edlbauer \textit{et al.} \cite{Edlbauer2021} replaced gold with lighter aluminum electrodes. Performing time-of-flight measurements, the authors demonstrated the generation of a strong and robust SAW confinement potential (above 24~meV peak-to-peak) which keeps the electron trapped in the same SAW minimum even in the presence of large variations (tens of meV) in the electrostatic potential landscape along the gated channel.

A similar aluminum IDT was employed later on by Wang \textit{et al.} \cite{Wang2023} and the enhanced transduction proved essential for synchronising two flying electrons, enabling the realisation of SAW-mediated collision experiments. Despite the improvements in transduction efficiency, it was still required to apply a strong radio-frequency input power to the IDT (above 25~dBm). The associated direct electromagnetic coupling disturbs the confinement potential at the source QDs before the arrival of the SAW \cite{phd_wang}. Furthermore, such a high input power also introduces unwanted heat into the system.

To reduce the input power while maintaining high transduction efficiency, one strategy is to employ unidirectional IDT designs \cite{Ekstrom2017, Dumur2019} (see Fig.~\ref{fig:idts}c,d). The asymmetry in the transducer's unit cell promotes the reflection of waves propagating in the backward direction \cite{Morgan2010} which superpose constructively with the waves propagating in the forward direction. Therefore, the unidirectional SAW emission provides an amplitude increase of 3~dB compared to regular IDTs.

From the material perspective, the weak electromechanical coupling of GaAs is the main limiting factor of the transduction efficiency. A solution is to deposit the IDT on a thin film of stronger piezoelectric substrates such as lithium niobate (LiNbO$_3$), aluminum nitride (AlN) or zinc oxide (ZnO). For instance, it has been shown that ZnO can be integrated with GaAs by depositing a metal film \cite{yuan2017} or a SiO$_2$ layer \cite{Dong2019} in-between. At last, circuit impedance matching should be considered to further enhance the acousto-electric amplitude.

\subsection{SAW engineering via Fourier synthesis}
\label{sec:fourier-synthesis}

Currently, the SAW employed for single-electron transport has typically hundreds of confinement locations. A major detrimental consequence of this long acoustic wave is that the unwanted minima can disturb the quantum state encoded in the electron \cite{Bertrand2016}. Furthermore, as explained in section~\ref{sec:single-electron-transport}, the control of the sending time relies on a picosecond voltage pulse applied on the source QD. This requires not only fine tuning of the QD potential, but also the need of one high-frequency line per single-electron source, rendering the triggering technique hard to scale. The solution to these problems is to engineer an acoustic wave with a single confinement potential.

\begin{figure}[!ht]
\centering
\includegraphics[width=80mm]{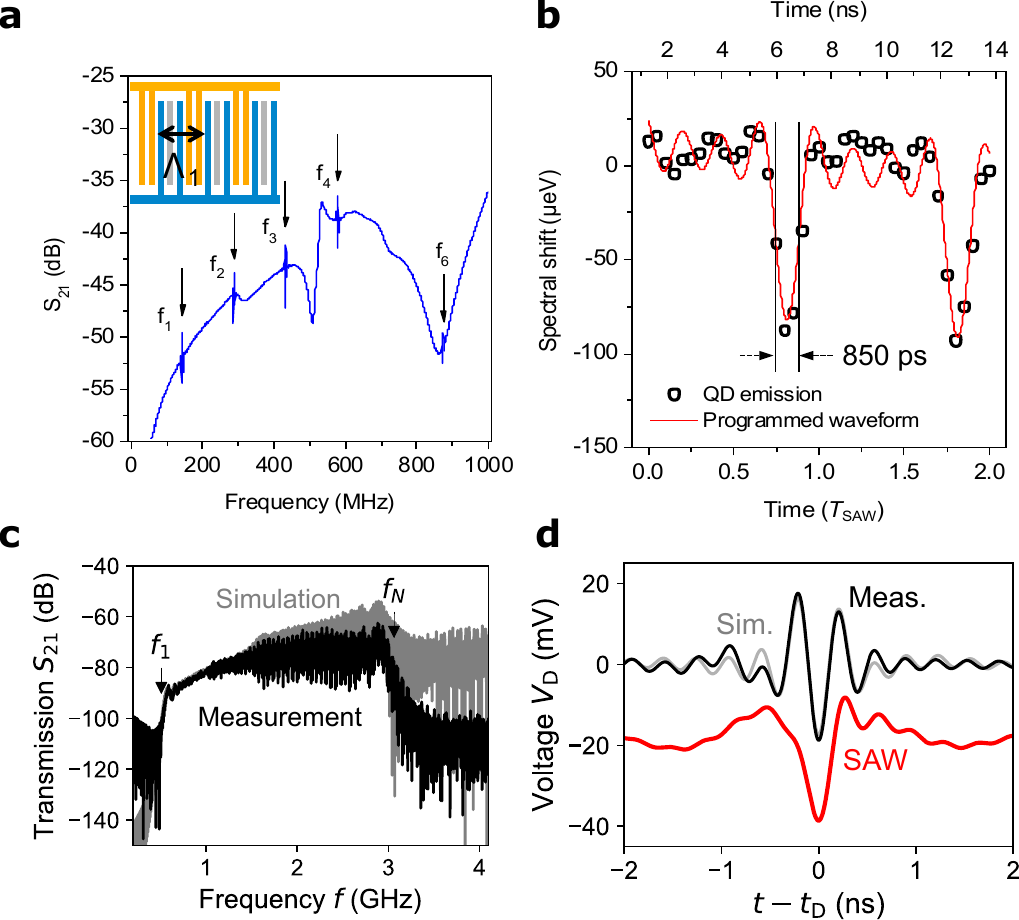}
\captionof{figure}{
    \textbf{SAW engineering.}
    \textbf{(a)} Frequency response of a Split-52 IDT featuring discrete harmonics, and 
    \textbf{(b)} the generated periodic delta-like SAW pulses.
    \textbf{(c)} Frequency response of a broadband chirp IDT, and 
    \textbf{(d)} the engineered single-cycle SAW pulse (red line).
    Figure adapted from \cite{Schulein2015} with permission from Springer Nature and \cite{Wang2022} with permission from the American Physical Society.
    \label{fig:saw_pulse}
}
\end{figure}

A pioneering SAW-engineering experiment was performed by Sch\"ulein \textit{et al.} \cite{Schulein2015}. The authors designed a novel transducer called Split-52 (see Fig.~\ref{fig:idts}e). This IDT has multiple harmonics from the fundamental resonance frequency (see Fig.~\ref{fig:saw_pulse}a). Based on Fourier synthesis, the authors excited different harmonics with carefully calibrated power and phase, and showcased the ability to engineer the acousto-electric shape. Figure~\ref{fig:saw_pulse}b shows the particular case of a periodic SAW train with delta-function pulses.

The generation of a single-cycle SAW pulse, \textit{i.e.} non-periodic, was demonstrated recently by Wang \textit{et al.} \cite{Wang2022}. Here, the authors employed a non-uniform transducer, the so-called chirp IDT \cite{Morgan2010} (see Fig.~\ref{fig:idts}f). Owing to its gradually-changing electrode periodicity, the frequency spectrum shows, in contrast to Split-52, a continuous broadband response (see Fig.~\ref{fig:saw_pulse}c). Based on the same principle of Fourier synthesis, an input chirp signal was applied to generate acoustic waves with gradually increasing frequencies. When these individual waves are perfectly synchronised, the interference results into a single delta-function-like acoustic pulse (see Fig.~\ref{fig:saw_pulse}d). The authors further validated the efficiency of such a single-cycle SAW pulse for single-electron transport between QDs, showing that the electron is transferred beyond 99~\% efficiency while being confined in the single confinement potential of the acoustic pulse.

Another advantage of SAW pulses compared to continuous SAWs is the suppression of the electromagnetic cross-talk, which is known as a limiting factor for pump accuracy \cite{Kataoka2006}. SAW pulses with well-defined sending intervals indeed allow to separate the arrival times of electromagnetic field and SAW pulse at the quantum channel. A pulse compression technique can be used to generate successive SAW pulses with controlled amplitudes and delays. More generally, arbitrary acoustic waveforms can be generated by simply changing the input radio-frequency signal. Finally, this novel SAW-engineering technique is not only limited to SAW-assisted transport, but is also directly applicable to a wide range of quantum acoustic applications \cite{Delsing2019, Kobayashi2017, Satzinger2018, Yokoi2020, Chen2021, Lyons2023}.

\subsection{On-demand single-electron source with a single-cycle SAW pulse}
\label{sec:saw_current}

A compressed SAW pulse with a single potential minimum has a couple of advantages for scaling up a qubit network. First, it can drastically simplify the synchronisation between multiple single-electron sources by removing the need for the pulse-trigger technique explained in Fig.~\ref{fig:trigger}. Second, it can also simplify the realisation of an on-demand single-electron source by removing completely the need for the source QD, with the SAW pulse directly picking up the single electron from the reservoir.

When a sinusoidal SAW -- generated by a standard IDT -- travels through a depleted quantum wire, a continuous flow of individual electrons is generated, each minimum of the sinusoidal potential being occupied by one electron \cite{Shilton1996}. This technique was first applied to metrology \cite{cunningham1999, cunningham2000, Janssen2000} and later to study the dynamics of individual flying electrons \cite{Kataoka2009, Ito2021}. Exploiting the chirp SAW pulse technique, it should be possible to realise an on-demand single-electron source since the timing of the SAW pulse can be arbitrarily controlled.

\begin{figure}[!ht]
\centering
\includegraphics[width=80mm]{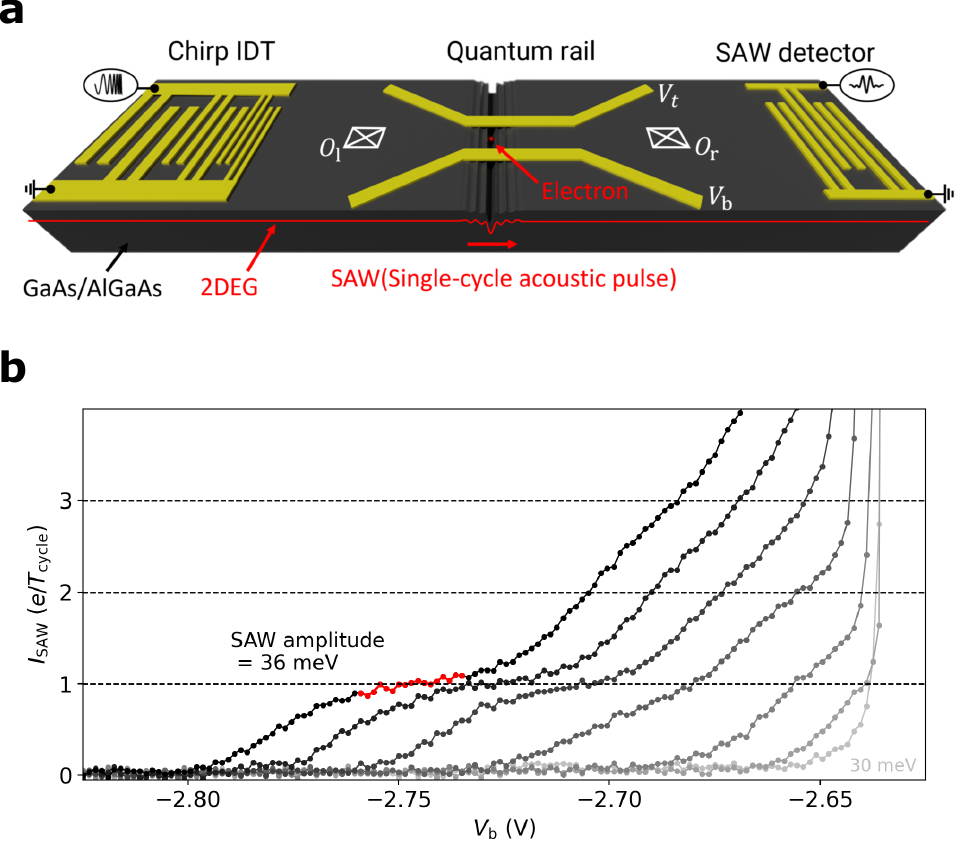}
\captionof{figure}{
    \textbf{On-demand single-electron source with a quantum wire and a single-cycle SAW pulse.}
    \textbf{(a)} Schematic of the device composed of a chirp IDT on the left, a quantum wire in the center and a SAW detector on the right.
    \textbf{(b)} Normalised acousto-electric current $I_{\rm SAW}$ induced by the single-cycle SAW pulses as a function of the voltage $V_{\rm b}$ at $V_{\rm t}=-2.2$~V. The SAW amplitude varies from 30 to 36~meV (from right to left). The range indicated in red corresponds to the flattest part where the SAW current is almost quantised.
    Figure adapted from \cite{ota2023ondemand} with permission from the American Physical Society.
    \label{fig:chirpQPC}
}
\end{figure}

Based on this idea, a single-electron pumping experiment with SAW pulses has been performed \cite{ota2023ondemand} using the device shown in Fig.~\ref{fig:chirpQPC}a. The central quantum wire was completely depleted by applying a large negative gate voltage to prevent electrons from flowing through the quantum wire when biasing the ohmic contact. An acousto-electric current $I_{\rm SAW}$ was then generated by applying SAW pulses with a repetition period $T_{\rm cycle}=1.28$~µs. In the event that each SAW pulse carries $n$ electrons of elementary charge $e$, the expected current is $I_{\rm SAW}=ne/T_{\rm cycle}$. Figure~\ref{fig:chirpQPC}b shows the normalised acousto-electric current as a function of the gate voltage on the quantum wire for different SAW amplitudes. For small amplitudes, the current smoothly decreases as the gate voltage is swept to more negative values. On the other hand, for larger amplitudes, the current shows a flatter dependence on the gate voltages around the value $e/T_{\rm cycle}$ corresponding to a single electron transferred by each SAW pulse.

In the flattest region indicated by the red points, the deviation of the current from an ideal single-electron pump is better than 4~\%. This value is however two orders of magnitude larger than achieved by the most accurate SAW-based single-electron pump whose deviation is about $10^{-4}$ \cite{Ford2017}. The employed chirp IDT was originally designed to generate delta-function SAW pulses, which turned out to be actually not the best choice to maximise transfer efficiency (see red dashed line in Fig.~1c of Ref.~\cite{ota2023ondemand}). To make the electron transport more accurate, besides improved IDT and pulse design, it will be also important to increase the SAW amplitude. For example, impedance matching of the chirp IDT to 50~$\Omega$ and improvement of the conversion efficiency between electromagnetic fields and acoustic waves by employing a piezoelectric thin film such as ZnO or AlN should be investigated.

Although the present accuracy at the $10^{-2}$ level is not yet useful for metrology purposes, the on-demand single-electron source developed in Ref.~\cite{ota2023ondemand} could be useful for single-electron quantum optics experiments using the charge degree of freedom. It does not require the electron preparation in a QD for electron sending, and hence can operate potentially faster. In addition, the number of single-electron sources can be increased by simply forming many quantum wires in parallel and transferring several single electrons at the same time owing to the wide wavefront of the SAW pulse. The synchronisation of all sources is also guaranteed by the single potential minimum of the SAW pulse. We anticipate accordingly, that synthesised acousto-electric pulses will strongly facilitate quantum experiments using single flying electrons transported by SAW.

\section{Single-electron to single-photon conversion}
\label{sec:single-photon-source}

Coupling between distant localised electron qubits for quantum computation or quantum communication requires the coherent conversion of an electron into a photon. The generation of a single photon from a single electron in a semiconductor nanostructure can be achieved by controlling the recombination of the single electron with a hole of the valence band. This challenging single-particle recombination process has recently been achieved by Hsiao \textit{et al.} \cite{Hsiao2020}.

\begin{figure}[!ht]
\centering
\includegraphics[width=88mm]{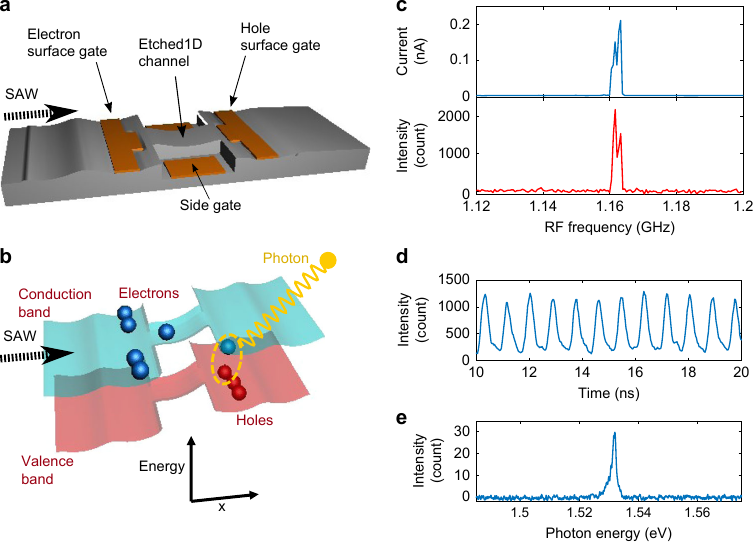}
\captionof{figure}{
    \textbf{Conversion from single electrons to single photons in a SAW-driven \textit{n-i-p} junction.}
    \textbf{(a)} Schematic of the device showing the surface gates used to induce electrons (\textit{n}-type region on the left) and holes (\textit{p}-type region on the right) in a GaAs quantum well, forming a lateral \textit{n-i-p} junction along an etched 1D channel. A SAW is generated by applying an RF signal to a transducer placed 1~mm from the junction.
    \textbf{(b)} Schematic diagram showing the band structure of the \textit{n-i-p} junction modulated by the SAW potential, for an applied forward bias less than the bandgap. A single electron is carried in each SAW minimum, creating a single photon when it recombines with a hole.
    \textbf{(c)} Source-drain current (top) and electro-luminescence intensity (bottom) as a function of applied RF frequency. Both signals show up around 1.163~GHz, which is the resonant SAW frequency of the IDT.
    \textbf{(d)} SAW-driven electro-luminescence intensity as a function of time. The 860~ps periodic feature corresponds to the applied SAW frequency of 1.163~GHz.
    \textbf{(e)} Energy spectrum of the SAW-driven electro-luminescence. The spectrum shows a peak at 1.531~eV which matches the exciton energy in the quantum well.  
    Figure reproduced from \cite{Hsiao2020} with permission from Springer Nature.
    \label{fig:photon-conversion}
}
\end{figure} 

In their experiment, a 2.5-µm-wavelength SAW is emitted from an interdigital transducer and then travels through a GaAs quantum well (QW) equipped with two surface gates in series to create a \textit{n-i-p} junction (Fig.~\ref{fig:photon-conversion}a). In the \textit{n}-type region where electrons are accumulated, the SAW splits the electron gas into wide stripes with several electrons in each potential well minima. In the intrinsic \textit{i} region between the two surface gates, the channel width is reduced by etching and by two side gates forming a quantum point contact (QPC) such that only one electron remains in each SAW minimum, the other electrons being reflected back into the \textit{n}-type region. Each single electron is then transported across the one-wavelength-long intrinsic region, where the 1.5~V potential barrier of the \textit{n-i-p} junction at equilibrium is reduced to a much smaller value using a large source-drain bias. In this regime, the slope of the electrostatic potential in the intrinsic region is small enough to not overcome the potential minima of the SAW. The single electrons can thus be efficiently transferred to the \textit{p}-type region of the junction, resulting in a nearly-quantised electrical current $I\sim e\,f_{\rm SAW}=0.186$~nA governed by the SAW frequency $f_{\rm SAW}\sim1.163$~GHz (Fig.~\ref{fig:photon-conversion}c, top panel). A detailed study of the current quantisation induced by SAW transport and QPC filtering has also been performed for \textit{n-n} and \textit{p-p} junctions \cite{Chung2019}.

For the \textit{n-i-p} junction case presented here \cite{Hsiao2020}, the single electrons transported by the SAW arrive in the \textit{p}-type region where holes are accumulated by the second surface gate. The recombination process of the single electron with one of the many holes of this region produces a 809~nm photon whose energy corresponds to the 1.53~eV band gap of the GaAs QW (Fig.~\ref{fig:photon-conversion}e). This electro-luminescence is collected with a lens focused on the \textit{p}-type region and coupled to an optical fiber for detection at room temperature using a single-photon avalanche photodiode (SPAD). The internal quantum efficiency of the photon emission has been estimated to be only about 2.5~\% despite the dedicated etching of the QW around the \textit{p}-type region to avoid electron loss outside the recombination region. This low efficiency could be the result of non-radiative recombination via surface states along the edges and to late radiative recombination outside the micron-size region of the collected light.

Single-photon emission has been demonstrated using a Hanbury-Brown and Twiss (HBT) interferometer to measure the probability to have two photons arriving at the same time. A clear photon antibunching effect has been observed with a suppression of the second-order correlation function below the 0.5 threshold value for single-photon emission. This regime could be obtained owing to a 94~ps recombination time which is shorter than the 860~ps period of the single-electron arrivals in the SAW train (Fig.~\ref{fig:photon-conversion}d).

To apply this single-electron-to-photon conversion to quantum information transfer, the next step will be to demonstrate the conversion of an electron spin into a circularly polarised photon. Such an experiment will require a source of spin-polarised electrons. This could be achieved via several strategies: spin injection with ferromagnetic contacts, magnetic focusing using a perpendicular magnetic field, harnessing the spin-orbit interaction of holes in the valence band, or by employing 2D materials such as monolayers of transition metal dichalcogenide with non-equivalent and spin-polarised K and K' valleys \cite{Kalameitsev2019, Nie2023}.

Another perspective could be to combine the two-electron interferometer presented in section~\ref{sec:charge-transport} with this single-electron-to-photon conversion technique based on \textit{n-i-p} junctions to produce entangled pairs of photons. The antibunching observed in the partitioning statistics of the two interacting electrons would be transferred to the two photons emitted at the \textit{n-i-p} junctions placed in the output channels of the interferometer. This effect, typical to fermionic statistics, would be unusual for photons, whose bosonic statistics usually produce a bunching effect.

In order to make this route relevant for applications, it will be important to enhance the recombination rate to reach high fidelity in the single-electron to single-photon conversion. A promising proposal by Wiele \textit{et al.}~\cite{Wiele1998} is to employ a quantum dot integrated inside an optical cavity. When a single electron-hole pair is trapped in the dot, the cavity stimulates the recombination process enhancing the efficiency of single-photon emission. Experimentally, a laser is employed to generate electrons and holes at a remote location, and a SAW is used to transport these carriers towards the quantum dot, which traps a single electron-hole pair and enhances its radiative recombination. The periodic refilling of the dot at the SAW frequency generates a regular single-photon pump. Although so far only the quantum-dot system without the cavity has been implemented, it revealed already strong antibunching in the emitted light, a clear evidence for the formation of a single-photon train \cite{Couto2009, Hernandez2011}. Interestingly, the SAW can be employed not only to inject the carriers into an individual emitter \cite{Voelk2012}, but also to regulate the injection process of the excitons into a single quantum dot \cite{Schulein2013}.

\section{Electrons surfing on superfluids}
\label{sec:superfluids}

At low temperatures, the surface of superfluid helium-4 resembles an exceptionally flawless platform, free from the imperfections typical of most other materials. When electrons are positioned near this surface, they are drawn to it and float several nanometres above the liquid, forming a distinctive two-dimensional electron system (2DES) boasting the highest electron mobility ever observed in condensed matter systems ($\mu > 10^8$~cm$^2$V$^{-1}$s$^{-1}$) \cite{shirahama1995surface}. This system has been an ideal playground for studying strong Coulomb interaction between electrons, in particular the Wigner crystal \cite{wigner1934interaction, grimes1979evidence, peeters1987electrons, andrei1988observation, Kono2010}.

Electrons in this system hold promise for quantum information science due to their anticipated long coherence time \cite{platzman1999quantum}. However, the presence of surface excitation, stemming from the liquid nature of helium, poses a significant constraint \cite{Godfrin1995, badrutdinov2020unidirectional}. This challenge was addressed recently by utilizing solid neon as a substrate in vacuum, effectively mitigating surface excitation \cite{schuster2010proposal, yang2016coupling, koolstra2019coupling}. Compared to liquid helium, solid neon has a superior surface rigidity which strongly suppresses decoherence mechanisms induced by surface excitation. This ultra-clean system has enabled the achievement of unprecedented charge qubit coherence times \cite{Zhou2022}, now reaching up to 100~µs \cite{Zhou2023}. It is notable that condensed noble-gas elements, whether liquid or solid, with positive electron affinity, uniquely retain electrons on a free surface in vacuum. In contrast, all other materials, even those electronically insulating and atomically smooth, possess negative electron affinity, often harboring charged contaminants or dangling bonds on the surface that can trap excess electrons at atomic to molecular scales \cite{zwanenburg2013silicon}.

\begin{figure}[!ht]
\centering
\includegraphics[width=88mm]{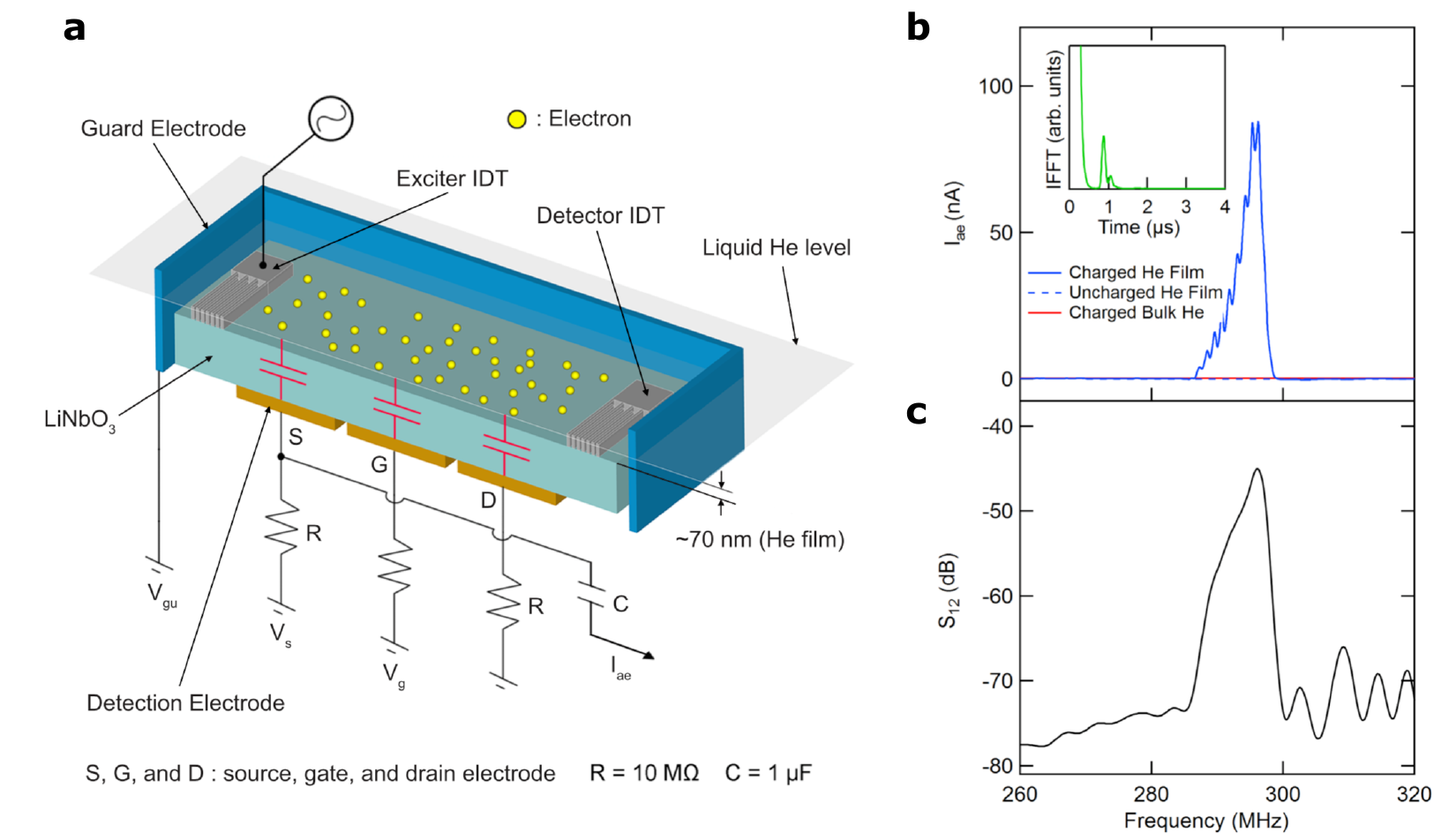}
\captionof{figure}{
    \textbf{Schematic of electron transport on top of helium using SAWs.}
    \textbf{(a)} Cross-section view of the device. Two opposing interdigital transducers (IDTs) are used to excite and receive SAWs. A saturated superfluid $^4$He film is formed on the surface of the LiNbO$_3$ piezo-substrate at 1.55~K. Electrons are trapped above the surface of the superfluid film by applying positive bias voltages to three underlying electrodes arranged in a field-effect transistor configuration. Lateral confinement of the electron layer is achieved with a negative bias to a guard electrode positioned on the outside of the LiNbO$_3$ substrate. 
    \textbf{(b)} Measured acousto-electric current $I_{\rm ae}$ of electrons on helium driven by the SAW as a function of frequency. Inset: Fourier transform of the signal which reveals a peak due to a SAW interference in the device. 
    \textbf{(c)} Frequency dependence of the transmission coefficient $S_{12}$ of the SAW device, demonstrating an expected resonance at 296~MHz. 
    Figure reproduced from \cite{byeon2021piezoacoustics} with permission from Springer Nature.
    \label{fig:e-on-He}
}
\end{figure} 

Helium-bound electrons offer promising prospects for the development of flying qubits. A recent study \cite{byeon2021piezoacoustics} demonstrated electron transport on the surface of helium using evanescent SAWs. The system, depicted in Fig.~\ref{fig:e-on-He}a, features electrons floating above a 70-nm-thick superfluid film, supported by a lithium niobate (LiNbO$_3$) piezoelectric substrate. By exciting a high-frequency acoustic wave, electrons are transported across the surface. This piezo-acoustic technique enables precise transportation of a small fraction of all electrons, down to about 0.01~\%, facilitating quantized charge pumping experiments. Besides single-electron transport, SAWs offer a pathway to directly investigate the high-frequency dynamic response and relaxation.

For the development of flying electron qubits with SAWs, it would be highly interesting to combine the approach discussed above for electrons on helium with the electron on neon system. If neon could be adsorbed on a piezoelectric substrate such as LiNbO$_3$ and if electron reservoirs could be engineered, one could envision single-electron transport in a system with minimal background perturbations, and thus coherent in-flight quantum manipulation of charge and spin degrees of freedom should be possible. 

For the semiconductor approach which we extensively discussed in this review, this is also possible by moving from doped semiconductor heterostructures to undoped heterostructures \cite{Kane1993, Harrell1999, Choi2020}. 
Such change would remove the random electrostatic charge background due to the ionised dopants in the heterostructure \cite{Nixon1990}. Although this comes with the drawback of a more complicated nanofabrication owing to the need of accumulations gates, it is certainly possible to realise, as experiments on spin qubits with Si/Ge \cite{Yoneda2017, Philips2022} and strained Ge \cite{Hendrickx2021, Scappucci2020} have shown over the last years.

\section{Conclusion}
\label{sec:conclusion}

Remarkable progress has been achieved over the last decade in the field of SAW-driven single-electron transport. The aim of the present review was to highlight some of the major milestones obtained in this field, with the focus on GaAs heterostructures.

First, we have seen that single-electron transfer probability has now reached values well above 99~\%, with transmission distances of over 60~µm, which makes this technology promising for applications in quantum technology. Single-electron couplers with similar precision have then been developed as first elements of future quantum gates.

Two different on-demand single-electron sources have been realised that allow to synchronise several single electrons flying in different electronic circuits. One is based on the triggering of the electron in a precisely-defined minimum of a long SAW train, the other exploits the engineering of a single SAW minimum in which the electron is carried.

Recently, these techniques have been used to collide two individual electrons on a beam splitter and record their partitioning statistics. In this experiment, the mutual Coulomb interaction between the two electrons was demonstrated for the first time by single-shot measurements of the electron antibunching. The interaction strength has proven to be extremely strong, so that in future experiments it should be possible to realise a controlled-phase gate with an interaction distance of well below one micrometer.

Continuing in this line, promising experiments on single-electron-to-photon conversion in the perspective of quantum information transfer have also been achieved. This novel conversion interface marks the first major step towards long-distance semiconductor qubit transfer via single optical photons.

Another important milestone was the demonstration of coherent spin transport over a macroscopic distance. In this experiment, entanglement over large distances was achieved between two electrons that were initially in a two-electron singlet state and were separated in a controlled manner over a distance of 6~µm. All along the displacement the electron spin undergoes coherent spin rotations and traces a new route towards fast on-chip deterministic interconnection of remote quantum bits in semiconductor quantum circuits.

Although these achievements were obtained using the simplest and well-established SAW generation technology (the regular IDT), we have pointed out that more sophisticated transducers and signal waveforms enable single-electron transport with better precision, synchronisation and scalability than the regular approach. In particular, implementations of chirp transducers producing acousto-electric pulses allow SAW-driven electron-quantum-optics experiments in GaAs devices without source and receiver QDs, strongly simplifying device geometries. Furthermore, there is plenty of room for enhancements, via unidirectional and focusing IDT designs, impedance matching and materials optimisation, for providing stronger in-flight confinement of the electrons within the SAW minima.

Beyond this traditional GaAs platform, we have discussed recent demonstrations of SAW-driven electron transport on the surface of superfluid helium and pointed out exciting routes for implementations on the surface of solid neon.

\section{Perspectives}
\label{sec:perspectives}

Let us now discuss several on-going investigations based on tunnel-coupled single-electron circuits for in-flight quantum manipulation of SAW-driven electrons. Compared to photons and electrons propagating in a Fermi sea, SAW electrons propagate very slowly, at the velocity of an acoustic vibration. This specificity enables their potential landscape to be varied in real time during the electron propagation, using the most recent radio-frequency equipment reaching 80~GHz bandwidth. By sending a properly engineered waveform on the tunnel-barrier gate of the circuit, the dynamical evolution of the electron state can be controlled during its propagation. Possible experiments include the investigation of energy relaxation and Landau-Zener transitions during the flight, the mapping of the static potential inhomogeneities along the channel, and the control of the coherent tunneling oscillations between the two channels.

Regarding the Coulomb-mediated coupling revealed and quantitatively measured via the antibunching effect in the two-electron collision experiment, further investigations can be carried out to better characterise the Coulomb interaction in moving quantum dots. In particular, more than two electrons can be launched simultaneously from the source QDs to study the partitioning statistics of a multi-electron interacting system \cite{Shaju2025}. The single-shot measurement capability of the experiment enables to record the complete set of partitioning probabilities that can be analysed using the full-counting statistics formalism. Such a study will provide quantitative information on the electrostatic interaction between two small sets of electrons flying in two adjacent tunnel-coupled channels. 

Another direction is the development of high-impedance SAW resonators employing focusing IDTs \cite{Lima2003} to create strongly confined resonant modes \cite{Msall2020}. These resonators are expected to exhibit large vacuum electric-field fluctuations and have the potential for strong capacitive coupling to a variety of solid-state quantum systems that couple to electric fields \cite{Kandel2024}. In the past, photon-assisted tunneling in a double quantum dot has been observed by coupling the dot to a standard traveling acousto-electric wave \cite{Naber2006}. Now, the coupling strength can be made large enough to dominate over both the typical dephasing rate of charge qubits and the decay rate of the piezoelectric resonator. This requires in particular cavities with high quality factors obtained by properly designing the resonator geometry. The delicate coupling between the phonon field and the qubit also requires a precise knowledge of the local strain field in the piezoelectric material. For this purpose, standing waves in IDT resonators can be characterised by direct imaging of the strain field using X-ray diffraction techniques \cite{Hanke2023}. Having achieved a strong enough coupling, the objective will be to couple the single charge of a quantum dot to a single phonon of the cavity, and demonstrate a coherent transfer of quantum information. On a longer term, a spin-phonon coupling might be achieved through a weak or intermediate spin-orbit coupling, but this would require an even stronger charge-phonon coupling. To achieve this ultimate goal, more sophisticated hybrid architectures might also be envisioned such as coupling the qubit to the high-impedance piezoelectric resonator through an even-higher-impedance superconducting resonator. These perspectives show the huge potential of SAWs for creating quantum interconnects between solid-state qubits.

\begin{acknowledgments}

We are very much indebted to Pierre-André Mortemousque for his valuable assistance in the preparation of the manuscript. This project has received funding from the European Union H2020 research and innovation program under grant agreement No. 862683, ``UltraFastNano''. C.B. acknowledges funding from the French Agence Nationale de la Recherche (ANR), project ANR QCONTROL ANR-18-JSTQ-0001 and project QUABS ANR-21-CE47-0013-01. C.B. and H.S. acknowledge funding from the Agence Nationale de la Recherche under the France 2030 programme, reference ANR-22-PETQ-0012. T.M. acknowledges funding from the Agence Nationale de la Recherche under the France 2030 programme, reference PEPR-PRESQUILE-ANR-22-PETQ-0002. J.W. acknowledges the European Union H2020 research and innovation program under the Marie Skłodowska-Curie grant agreement No. 754303. S.T. acknowledges financial support from JSPS KAKENHI grant No. 20H02559, 23H00257 and JST Moonshot R\&D grant No. JPMJMS226B.

\end{acknowledgments}

\bibliography{references}

\end{document}